\documentclass[%
reprint,
superscriptaddress,
 amsmath,amssymb,
 aps,
 prx,
]{revtex4-1}

\usepackage{graphicx}
\usepackage{dcolumn}
\usepackage{bm}
\usepackage[dvipsnames]{xcolor}

\usepackage[ruled,vlined]{algorithm2e}
\usepackage{algorithmic}
\usepackage[normalem]{ulem}
\usepackage{url}

\usepackage[utf8]{inputenc}
\usepackage[T1]{fontenc}

\newtheorem{theorem}{Theorem}

\begin{document}

\preprint{APS/123-QED}

\title{One-step replica symmetry breaking in the language of tensor networks}

\author{Nicola Pancotti}
\affiliation{AWS Center for Quantum Computing, Pasadena, CA 91125, USA}
\affiliation{California Institute of Technology, Pasadena, CA, USA}
\author{Johnnie Gray}
\affiliation{Division of Chemistry and Chemical Engineering, California Institute of Technology, Pasadena, USA 91125}

\date{\today}

\begin{abstract}

We develop an exact mapping between the one-step replica symmetry breaking cavity method and tensor networks.
The two schemes come with complementary mathematical and numerical toolboxes that could be leveraged to improve the respective states of the art.
As an example, we construct a tensor-network representation of Survey Propagation, one of the best deterministic {\it k}-SAT solvers.
The resulting algorithm outperforms any existent tensor-network solver by several orders of magnitude.
We comment on the generality of these ideas, and we show how to extend them to the context of quantum tensor networks.
\end{abstract}

\maketitle

\section{Introduction}\label{sec:intro}

Tensor networks have recently emerged as a powerful technique to study high-dimensional problems.
In recent years, they enjoyed extensive applications in various disciplines including multilinear algebra, optimization, differential equations and machine learning.
At the same time, their intrinsic linear structure admits sophisticated mathematical reasoning that has led to numerous results of practical and theoretical interest~\cite{cichocki2016tensor, cichocki2017tensor, cirac2021matrix, orus2014practical, evenbly2022practical}.
We are currently witnessing an ever growing characterization of small and regular tensor networks, like MPS~\cite{perez2006matrix, oseledets2011tensor}, PEPS~\cite{verstraete2004renormalization}, and MERA~\cite{vidal2007entanglement}.
However, despite some attempts to numerically study large random networks~\cite{gray2021hyper, gray2022hyper}, we still lack a robust theory and algorithms capable of assessing the hardness of exact or approximate contractions.

Statistical mechanics of disordered systems specializes in the opposite situation where the number of degrees of freedom is large and the corresponding graph is random and non local.
In this context, the theory of Replica Symmetry Breaking (RSB)~\cite{mezard1987spin, nishimori2001statistical, altieri2023introduction} and its algorithmic extensions was invented to tackle problems beyond naive mean-field strategies.
Quite interestingly, RSB comes equipped with a robust mathematical theory~\cite{talagrand2010mean} that produced a number of exact and rigorous results in physics and computer science~\cite{mezard1987spin, nishimori2001statistical, Mezard_Montanari_book_2009}.
Although it was originally developed in the context of disordered systems and spin glasses, in the last forty years the RSB theory has emerged as a universal language to reason about inference, classical information processing and combinatorial optimization~\cite{nishimori2001statistical, Mezard_Montanari_book_2009, zdeborova2016statistical, mezard2002analytic, montanari2021optimization}.
In particular, the one-step Replica Symmetry Breaking (1RSB) cavity method~\cite{mezard_bethe_2001, mezard_cavity_2003} in its formulation in terms of factor graphs can be identified as the one framework that encompasses many of the most interesting practical applications.

Here we build on the well-known correspondence between tensor networks and factor graphs~\cite{robeva2019duality} to develop a formulation of the 1RSB cavity method in the language of tensor networks.

Although any factor graph admits an exact representation as a tensor network, the opposite is not necessarily true.
However, we will show that, up to an overall summation, tensor networks can interpreted as factor graphs.
This allows us to employ Belief Propagation (BP)~\cite{pearl1988probabilistic} to extend the results in Ref.~\cite{Alkabetz2021} and expand the mean-field theory of tensor networks.
In particular, we show how to associate conventional statistical objects, e.g. the free entropy, to a random tensor network, and how to express them in terms of local quantities like marginals and BP fixed-points.
Interestingly, this establishes a correspondence between the variational optimization of the free entropy and the full tensor network contraction.

Many classical results can be reformulated within this framework.
Here, we focus on the 1RSB cavity method.
By following the recipe in Ref.~\cite{Mezard_Montanari_book_2009}, we show that, given a random tensor network on a loopy graph, it is possible to construct an auxiliary tensor network in a similar fashion to factor graphs.
This auxiliary model is the central object of study as it has been shown that several features beyond mean-field can be efficiently learned by analyzing it.

To showcase a practical example, we consider combinatorial optimization as one of the most celebrated applications of the 1RSB cavity method~\cite{mezard2002analytic, mezard2002random, braunstein2005survey, Maneva2007, krzakala2007gibbs, marino2016backtracking}.
In a series of ground-breaking results, physicists developed sophisticated statistical theories of many problems in this class.
These include the maximally independent set (MIS)~\cite{barbier2013hard}, the assignment and traveling salesman problem~\cite{mezard1986mean}, graph coloring~\cite{mulet2002coloring, braunstein2003polynomial}, low density parity check (LDPC) codes~\cite{sourlas1989spin, montanari2001finite, montanari2001glassy}, and constraint satisfaction problems~\cite{mezard2002analytic, mezard2002random, braunstein2005survey}.
The 1RSB cavity method, besides giving a refined description of the problem, was shown to be exceptionally efficient in finding solutions close to the complexity phase transitions.

There have been many alternative attempts to solve combinatorial optimization problems with physics-inspired techniques.
Notably, Monte Carlo methods~\cite{metropolis1949monte, hastings1970monte} have been extensively employed as gradient-free methods, demonstrating remarkable performance~\cite{kirkpatrick1983optimization, earl2005parallel, zhu2015efficient, wang2015comparing, barzegar2018optimization, barzegar2021optimization}.
A thorough comparison between modern (non-equilibrium) Monte Carlo algorithms and the 1RSB cavity methods was performed in Ref.~\cite{mohseni2021nonequilibrium}.
The authors provide a number of benchmarks that demonstrate significant speedup against both specialized 1RSB algorithms and generic stochastic solvers.

Physics-inspired machine learning methods, on the other hand, were recently proposed as a very attractive alternative to attack this kind of problems on random graphs.
Despite their infancy, they have already displayed noteworthy achievements for MIS and graph coloring~\cite{schuetz2022combinatorial, angelini2022modern, schuetz2022reply, schuetz2022graph}.

Combinatorial optimization has attracted substantial interest from the quantum community as well.
In the recent years there have been several proposal to use quantum devices to solve complex combinatorial tasks.
These include theoretical proposals~\cite{albash2018adiabatic, farhi2001quantum, mcclean2021low, mandra2016strengths, mandra2016strengths, karimi2017effective, mandra2018deceptive, aramon2019physics} and experimental demonstrations~\cite{pichler2018quantum, zhou2020quantum, ebadi2022quantum}.
Given the current limitations of quantum hardware, it is customary to employ tensor-network techniques to test and benchmark their performance.

In recent years, tensor networks have been also proposed as a standalone approach~\cite{kourtis2019fast, kissinger2020tensor, Rams2021, liu2021tropical, liu2022computing}.
Their exceptional power in characterizing correlated quantum systems, strongly motivates their use in the context of highly correlated classical problems.
However, despite several attempts, we currently lack algorithms that compete with state-of-the-art heuristic solvers.
The main drawback of current approaches is that they often solely rely on the geometrical properties of the graph, and do not explicitly exploit the structure of the correlations.
Therefore, they easily incur exponential costs.

Here, we propose an alternative route which is inspired from the 1RSB cavity method.
We map Survey Propagation (SP)~\cite{braunstein2005survey}, one of the best deterministic {\it k}-SAT solvers, to tensor networks.
This allows us to solve {\it k}-SAT instances with thousands of variables in the vicinity of the SAT-UNSAT transition.
Our mapping is very general and can be used to translate many of the classical algorithms based on message-passing techniques into the tensor-network framework.
As most of these algorithms often rely on mean-field routines like BP, it is conceivable that more advanced contraction strategies will be able to improve their performance~\cite{wang2023tensor}.

The range of applicability is not limited to classical problems.
We show how to extend these ideas to tensor networks describing quantum wavefunctions.
Despite containing complex numbers, we show that their associated 1RSB auxiliary model is always real and non-negative.
Namely, the auxiliary model is a classical distribution.

Our results open up a plethora of possibilities to leverage powerful mathematical and numerical toolboxes that have the potential to advance the mean-field theory of classical and quantum tensor networks and, at the same time, to enhance the classical theory of strongly correlated systems.

The paper is organized as follows.
In Sec.~\ref{sec:correspondence}, we outline the mapping between factor graphs and tensor networks.
We detail the theory of belief propagation for tensor networks and we derive some relevant quantities in terms of fixed points, such as marginals and free entropy.
In Sec.~\ref{sec:1rsb}, we survey the classical theory of the 1RSB cavity method and we develop its tensor network formulation.
In Sec.~\ref{sec:ksat}, we show how to use these techniques to find solutions of {\it k}-SAT instances.
In particular, we give a constructive algorithm to build the tensor network associated to Survey Propagation.
Finally, in Sec.~\ref{sec:quantum}, we derive the relevant quantities to build the auxiliary model when the tensor network describes a quantum wavefunction.

\section{Correspondence between tensor networks and factor graphs}\label{sec:correspondence}

\begin{figure}
\includegraphics[width=0.49\textwidth]{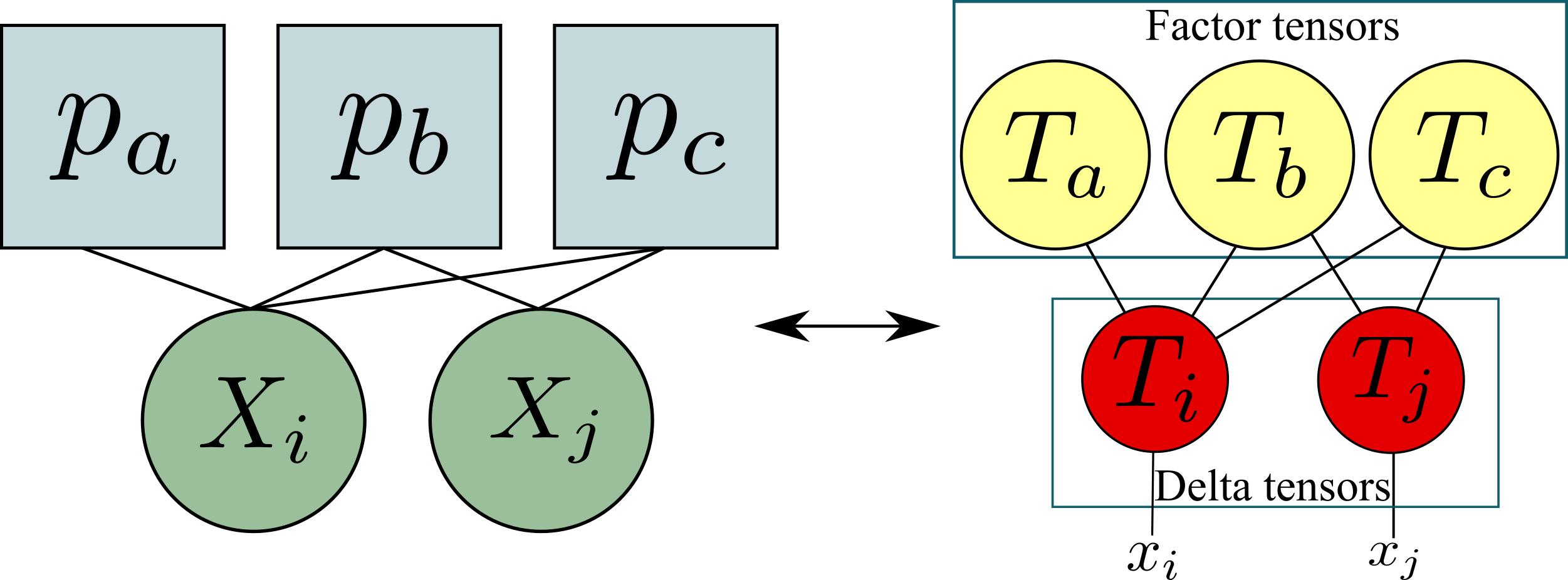}
\caption{{\bf Graphical representations. Left:} Graphical representation of the factor graph $p_a(X_i)p_b(X_i, X_j)p_c(X_i, X_j)$. Factor graphs can be always represented as a bipartite graph of variables (green circles) and factors (gray squares). {\bf Right:} Graphical representation of the corresponding tensor network. The red circles represent delta (or copy) tensors, whereas the yellow circles represent the tensors $\{T_a, T_b, T_c\}$, isomorphic to the factors $\{ p_a, p_b, p_c \}$. The indices $x_i$ and $x_j$ enumerate the values of $X_i$ and $X_j$, respectively.}
\label{fig:fg2tn}
\end{figure}

In this work, we consider probability distributions $P (X_1, X_2, \ldots, X_N)$ which factorize according to
\begin{equation}\label{eq:factor_graph_tn_section}
  P (X_1, X_2, \ldots, X_N) = \frac{1}{Z} \prod_{a \in F} p_a (X_{\partial a}), 
\end{equation}
where the variables $X_i \in \mathcal{X}$ take value in a finite discrete alphabet $\mathcal{X}$ (possibly variable dependent), $F$ is the set of all factors $p_a$, and $Z$ is the partition function.
For the moment we will assume $p_a \geq 0$, in Sec.~\ref{sec:quantum} we will relax that assumption.
With $\partial a$ we indicate the set of variables appearing in $p_a$. 
Probability distributions that admit this kind of factorization are usually referred to as {\it factor graphs} or {\it graphical models}.
Factor graphs can be represented graphically by introducing a bipartite graph where each vertex represents either a factor or a variable, and each edge connects a variable $X_i$ and a factor $p_a$ only if $X_i \in \partial a$.
In Fig.~\ref{fig:fg2tn} (left), we show the graphical representation of a simple factor graph.

We call a tensor a multilinear array with an arbitrary number of dimensions or indices.
We will denote vectors as $T[y]$ (one index), matrices as $T[y,z]$ (two indices), and so on.
The square brackets stress that the $T[\cdot]$'s are tensors, and that the variables between the square brackets are indices.
Clearly, any tensor can be interpreted as a function defined over a discrete domain.

Tensor networks are collections of tensors that share indices among each other.
They admit a natural graph representation where each vertex corresponds to a tensor, and an edge connects two (or more) vertices only if the corresponding tensors share an index.
Tensor networks can be very expressive and they have been widely used in quantum physics.
One would typically consider functions of the form
\begin{align} \label{eq:init_tn}
  P (\{x\}) \propto \sum_{\{r\}} \prod_{a \in F} T_a[x_{\partial a}, r_{\partial a}]
\end{align}
where we denoted by $\{x\}$ the set of indices that appear only once, and with $\{r\}$ more than once.
Sometimes, following Einstein's notation, the summation in Eq.~\eqref{eq:init_tn} is omitted.

Factor graphs and tensor networks share many similarities.
In this work, we will leverage their relationship to map data structures and algorithms back and forth between the two frameworks.

A tensor-network representation of $P (X_1, X_2, \ldots, X_N)$ in Eq.~\eqref{eq:factor_graph_tn_section}, is a collection of tensors such that, given any value of the variables $\mathbf{X}^* = (X^*_1, \ldots, X^*_N)$, the number $\prod_{a \in F} p_a (X^*_{\partial a})$ can be computed as a multilinear tensor contraction, similar to Eq.~\eqref{eq:init_tn}.
Since we are dealing with variables on a discrete alphabet, let us introduce an index $x_i$ that enumerates each distinct value of $X_i$.
There exists a simple isomorphism between tensors and factors. 
Because each variable is defined on a discrete alphabet, each factor $p_a (X_{\partial a})$ can take at most $|\mathcal{X}|^{|\partial a|}$ different values.
We can therefore define a tensor $T_a[x_{\partial a}]$ that, for each set of values $X^*_{\partial a}$, has a set of indices $x^*_{\partial a}$ such that $p_a (X^*_{\partial a}) = T_a[x^*_{\partial a}]$.
Using these definitions we can map any factor graph onto a tensor network. To each variable $X_i$ we associate a ``delta'' tensor $T_i [x_i, r_{\partial a}]$ with $| \partial i | + 1$ indices such that $T_i [x_i, r_{\partial a}] = 1$ only if $x_i = r_{k}$ for any $k \in \partial a$, and zero otherwise.
On the other hand, to each factor we associate a tensor $T_a [r_{\partial a}]$ with $| \partial a |$ indices that, for any value $r^*_{\partial a}$, $T_a[r^*_{\partial a}] = p_a (X^*_{\partial a})$.

In Fig.~\ref{fig:fg2tn}, we provide a graphical representation of this mapping for a simple factor graph of the form $p_a(X_i)p_b(X_i, X_j)p_c(X_i, X_j)$. Notice that the two graphs have the same topology, and each vertex of one graph is mapped to exactly one vertex of the other.

Eq.~\eqref{eq:factor_graph_tn_section} can be thus retrieved as
\begin{equation}\label{eq:tn_factor_graph}
    P (X_1, X_2, \ldots, X_N) \propto \sum_{\{r\}} \prod_{a \in F} T_a [r_{\partial a}] \prod_{i \in V} T_i [x_i, r_{\partial i}]
\end{equation}
where $V$ is the set of all variables.
Notice that the summation in Eq.~\eqref{eq:tn_factor_graph} can be easily carried out since the $T_i$'s are delta tensors, by definition.
Each of them splits into a product of independent terms which correspond exactly to Eq.~\eqref{eq:factor_graph_tn_section}, up to an overall factor.

The opposite is not necessarily true.
Not all tensor networks admit a (simple) factor graph representation.
However, any tensor network can be represented in the language of factor graphs as follows.
A typical tensor network can be defined as
\begin{equation}
  \label{eq:general_tn}
  \sum_{\{r\}} \prod_{k} T_k [x_k, r_{\partial k}].
\end{equation}
Notice that Eq.~\eqref{eq:general_tn} is nothing but a summation over a product of tensors $\prod_{k} T_k [x_k, r_{\partial k}]$ that can be as well interpreted as a factor graph.
Thus, if each $T_k$ has only real and positive entries, a tensor network can be seen as a marginalization of a regular factor graph.
Later we will deal with the case of complex numbers.
It is important to stress that a tensor network can sometimes reduce to a factor graph without marginalization.
The conditions such that this is true are not trivial, and they are not in the scope of this work.
One of our main goals is to shed light upon certain algorithms originally developed in the factor graph language in order to approximate marginal distributions.
These algorithms could be easily translated into the language of tensor networks.
Therefore, in some cases, we expect them to be able to handle the marginalization in Eq.~\eqref{eq:general_tn}.

\subsection{Belief propagation and tensor networks}

In Ref.~\cite{Alkabetz2021}, it was shown that there exists a deep connection between belief propagation (BP) and tensor network contractions.
The authors considered the problem of finding marginals of quantum wave functions (i.e., reduced density matrices).
The task comes with the complication of quantum probabilities, L2 norms, and so on.
Here we first consider the simpler scenario where each entry in the tensors is real and positive.
Notice that this condition is satisfied whenever we map a factor graph onto a tensor network as above.
We can therefore normalize our distribution with the usual L1 norm and directly compare the ``classical'' belief propagation to its tensor network counterpart.

Belief Propagation (BP) is an iterative algorithm that was designed to compute marginal distributions over factor graphs or graphical models.
The update BP equations for the factor graph in Eq.~\eqref{eq:factor_graph_tn_section} read,
\begin{align}
  \label{eq:BP_first}
  \nu_{j \rightarrow a}^{(t+1)} (X_j) & \cong \prod_{b \in \partial j \backslash a} \hat{\nu}_{b \rightarrow j} ^{(t)} (X_j), \\
  \label{eq:BP_second}
    \hat{\nu}_{a \rightarrow j}^{(t+1)} (X_j) & \cong \sum_{x_{\partial a \backslash j}} p_a (X_{\partial a }) \prod_{k \in \partial a \backslash j} \nu_{k \rightarrow a} ^{(t)} (X_k),
\end{align}
where we introduced the set of messages $\{\nu, \hat{\nu}\}$, which are functions of the variables $\nu, \hat{\nu}: \mathcal{X} \rightarrow \mathbb{R}_+$.
The goal of belief propagation is to iterate Eqs.~(\ref{eq:BP_first}, \ref{eq:BP_second}) until the left-hand-side messages are approximately equal to the right-hand-side and then use the resulting set of fixed-point messages $\{ \nu^*, \hat{\nu}^* \}$ to study the original problem.
If the factor graph is supported on a tree, the algorithm is guaranteed to converge to an unique fixed point, independent of initialization.
Namely, BP is exact on trees.
Many important statistical quantities take a close form in terms of messages.
In particular, free energy and free entropy, internal energy, marginals, and conditional distributions can all be computed efficiently from the fixed-point messages.
Additionally, once we have access to marginal and conditional distributions, we can efficiently sample from the factor graph.
These are some of the reasons why it is interesting to map belief propagation to its tensor-network counterpart.

By using the decomposition in Eq.~\eqref{eq:tn_factor_graph}, we can translate each one of the BP equations to a simple tensor contraction.
Let us introduce the environments $E^{(t)}_{i \rightarrow a} [r_{i_a}]$ and $E^{(t)}_{a \rightarrow i} [r_{a_i}]$ from variable tensors to factor tensors, and viceversa.
Each pair of environments can be understood as $\mathcal{X}-$dimensional vectors living in the bond between variable $i$ and factor $a$.

In what follows, the environments will play the same role of the messages and, in practice, they will carry the exact same information.
The only qualitative difference between messages and environments is that the latter provides an algebraic interpretation of the problem.
This allows to employ the rich collection of tools that have been developed in the context of tensor networks such as tensor-network contractions, decompositions, and compression, in the context of belief propagation and (linear) message-passing algorithms.
On the other hand, this also allows to leverage the robust theory behind belief propagation (and its extensions) to the theory of tensor networks.

The first BP equation can be rewritten as follows
\begin{equation}
  \label{eq:first_tnbp_eq_withvar}
  E^{(t+1)}_{j \rightarrow a} [r_{j_a}] = \sum_{x_j, r_{\partial j \backslash j_a}} T_j [x_j, r_{\partial j}]  \bigotimes_{b \in \partial j \backslash a} E^{(t)}_{b \rightarrow j} [r_{b_j}].
\end{equation}
To see how Eq.~\eqref{eq:first_tnbp_eq_withvar} reduces to Eq.~\eqref{eq:BP_first}, recall that $T_j [x_j, r_{\partial j}]$ is a delta tensor.
Thus, the summation reduces to only $| \mathcal{X} |$ terms of form $E^{(t+1)}_{j \rightarrow a} [r_{j_a}] = \prod_{b \in \partial j \backslash a} E^{(t)}_{b \rightarrow j} [r_{b_j}]$ which are equivalent to Eq.~\eqref{eq:BP_first} (the delta tensor enforces $x_j = r_{b_j}$ for all $b_j$).
Furthermore, notice that the variable $x_j$ does not appear in the environments.
We can thus absorb the corresponding sum as
\begin{equation}
  \label{eq:first_tnbp_eq}
  E^{(t+1)}_{j \rightarrow a} [r_{j_a}] = \sum_{r_{\partial j \backslash j_a}} T_j [r_{\partial j}] \bigotimes_{b \in \partial j \backslash a} E^{(t)}_{b \rightarrow j} [r_{b_j}],
\end{equation}
where $T_j [r_{\partial j}] = \sum_{x_j} T_j [x_j, r_{\partial j}]$.

\begin{figure}
\includegraphics[width=0.35\textwidth]{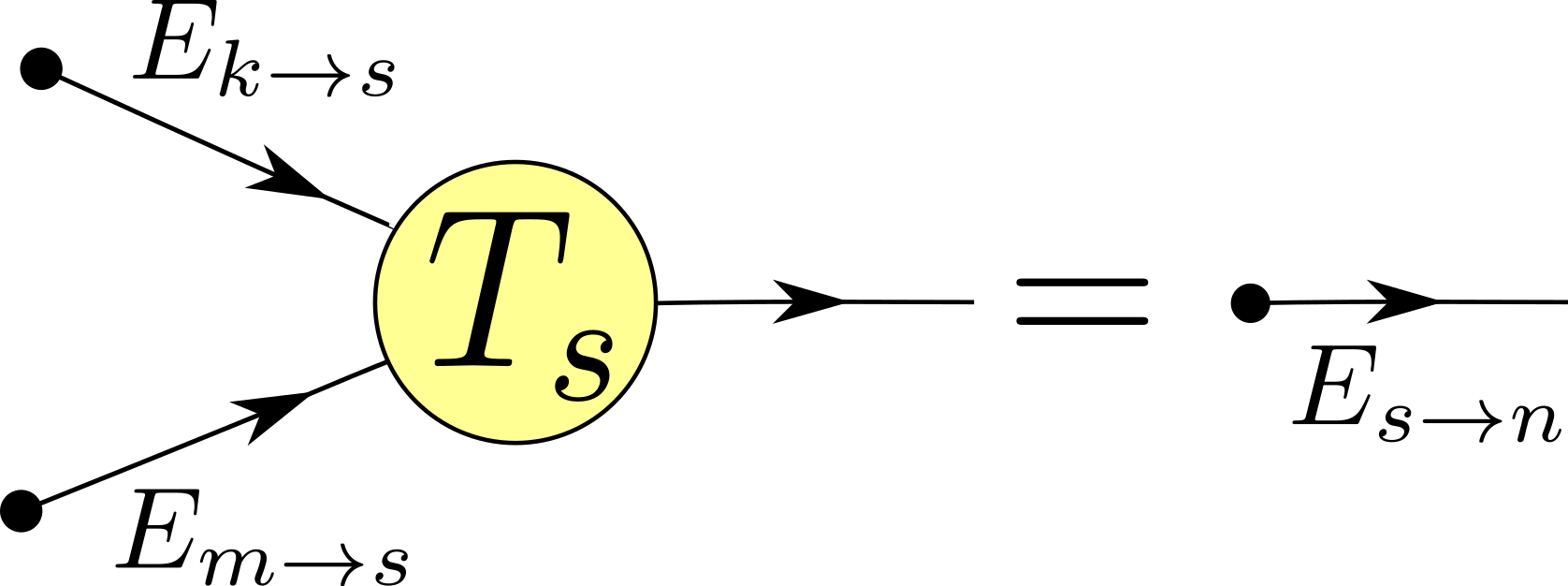}
\caption{{\bf Graphical representation of fixed point conditions in Eq.~\eqref{eq:fixed_point_tnbp}.} A fixed point of belief propagation is a set of environments $\{ E_{i \rightarrow j} \}$ that (approximately) satisfy the set of all equations of this form.}
\label{fig:tn_fxpnt}
\end{figure}

Similarly, the second BP equation can be written as
\begin{equation}
  \label{eq:second_tnbp_eq}
  E^{(t+1)}_{a \rightarrow j} [r_{a_j}] = \sum_{r_{\partial a \backslash j_a}} T_a [r_{\partial a}]  \bigotimes_{k \in \partial a \backslash j} E^{(t)}_{k \rightarrow a} [r_{a_k}],
\end{equation}
where the tensor product plays the role of the product in Eq.~\eqref{eq:BP_second}, and $T_a [r_{\partial a}]$ is isomorphic to $p_a (X_{\partial a})$, by definition.
Notice that the two tensorized BP Eqs.~(\ref{eq:first_tnbp_eq}, \ref{eq:second_tnbp_eq}) have exactly the same structure.
For simplicity, let us group them in a single equation in its fixed-point form
\begin{equation}
  \label{eq:fixed_point_tnbp}
  E_{s \rightarrow n} [r_{s_n}] = \sum_{r_{\partial s \backslash n_s}} T_s [r_{\partial s}]  \bigotimes_{k \in \partial s \backslash n} E_{k \rightarrow s} [r_{s_k}],
\end{equation}
such that when $T_s$ is a factor tensor, $T_n$ is a variable tensor, and viceversa.
In Fig.~\ref{fig:tn_fxpnt}, we provide a graphical representation of Eq.~\eqref{eq:fixed_point_tnbp} for a hypothetical tensor $T_s$ of rank three.

Note that Eq.~\eqref{eq:fixed_point_tnbp} defines a local linear map from the space of environments to itself. The fixed-point condition is the solution of a coupled linear system of equations.

\subsection{Efficient numerical implementation}\label{sec:vtnbp}

The tensorized formulation of belief propagation lends itself to very simple and efficient numerical implementations.
For simplicity, let us assume that the tensor network is supported on a regular graph of degree three.
This is not a very typical scenario, but it will simplify the exposition.
The generalization to arbitrary graphs is straightforward.
If all tensors $T_s[r_{s_v}, r_{s_w}, r_{s_n}]$ have the same number of indices, we can stack them together into a single tensor $\tilde{T} [s, r_{s_v}, r_{s_w}, r_{s_n}] = T_s[r_{s_v}, r_{s_w}, r_{s_n}]$.
Similarly, we can stack the environments $E_{s \rightarrow n} [s, r_{s_n}] = E_{s \rightarrow n} [r_{s_n}]$.
With these new data structures, the BP update can be carried out efficiently with a small number of contractions.
For instance, for a regular graph of degree three, a BP update of the whole tensor network requires three contractions in total.
Each one of those taking the simple form
\begin{equation}
  \label{eq:vtnbp}
  E_{s \rightarrow n} [s, r_{s_n}] = \sum_{r_{\partial s \backslash n_s}} \tilde{T} [s, r_{\partial s}]  \bigotimes_{k \in \partial s \backslash n} E_{k \rightarrow s} [s, r_{s_k}].
\end{equation}
After each contraction, the output environments need to be permuted back to the correct order.
This permutation enforces the original topology of the graph.
If the graph is not regular, it is necessary to simply group the tensors by degree and stack them accordingly.

Such a `vectorized' approach makes the updates amenable to various acceleration techniques developed for large contractions, and is particularly powerful for large graphs with only a few different tensor ranks.
~
As an example, a single BP update with this approach for a 
$50 \times 50 \times 50$ cubic lattice takes about a tenth of a second on a single CPU, with a similar number for {\it k}-SAT instances of 100,000 variables close to the complexity phase transition point.
\begin{figure*}
  \includegraphics[width=1.0\textwidth]{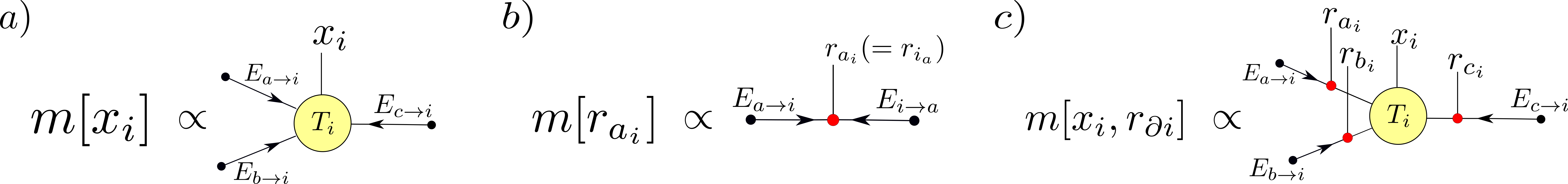}
\caption{{\bf Graphical representation of the marginal distributions.} Graphical representation of the single variable and local marginal distributions as tensor networks. The red dots represent copy tensors of the relevant variables. {\bf a)} Single visible/physical variable as in Eq.~\eqref{eq:local_marginals_tensor_network_x}. {\bf b)} Single bond variables as in Eq.~\eqref{eq:local_marginals_tensor_network_r}. {\bf c)} Marginal of a region around a variable as in Eq.~\eqref{eq:marginal_region_envs}.}
\label{fig:tn_marginals}
\end{figure*}

\subsection{Tensor factor graphs on trees}\label{sec:tensor_factor_graph}

In this section we consider tensor networks of the following form 
\begin{equation}
  \label{eq:tensor_factor_graph}
  T [\mathbf{x}, \mathbf{r}] = \prod_{i \in \mathcal{T}} T_i [x_i, r_{\partial i}],
\end{equation}
where $\mathcal{T}$ is the set of all tensors.
In order to develop the basic theory, we will first assume that the tensor network is supported on a tree graph.
This class has been extensively studied in the context of quantum mechanics~\cite{fannes1992ground, friedman1997density, lepetit2000density, shi2006classical, tagliacozzo2009simulation, nagaj2008quantum, murg2010simulating, li2012efficient, nakatani2013efficient, pivzorn2013tree, gerster2014unconstrained, murg2015tree}, here we outline a deep connection with the theory of factor graphs and disordered systems.
If we relax the condition that each tensor $T_i [x_i, r_{\partial i}]$ must depend on $x_i$, it is easy to see that Eq.~\eqref{eq:tensor_factor_graph} encompass the factor graph decomposition from the previous section.
At the same time, Eq.~\eqref{eq:tensor_factor_graph} can be also interpreted as a factor graph itself with two disjoint sets of variables: $x_i$ and $r_j$.
The peculiar feature of this class is that each $x_i$ appears in one term only, whereas each $r_j$ appears in two terms only.
We will refer to this class of factor graphs as {\it tensor factor graphs} since any factor graph, once it is mapped to the corresponding tensor network, can be represented this way.
And, any tensor network as in Eq.~\eqref{eq:general_tn} also admits the same description up to an overall summation.

As it was noticed in Ref.~\cite{Alkabetz2021}, because each index $x_i$ appears only once per tensor, it is always possible to efficiently carry out the corresponding summation without affecting the topology of the graph.
Indeed, the tensor network 
\begin{equation}
  \label{eq:marginal_visible}
  \sum_{\mathbf{x}} T[\mathbf{x}, \mathbf{r}] = \prod_{i \in V} T_{i} [r_{\partial i}] 
\end{equation}
can be expressed as a product of terms where the nodes corresponding to the indices $\mathbf{x}$ have been absorbed as $T_{i} [r_{\partial i}]  = \sum_{x_i} T_{i} [x_i, r_{\partial i}]$.
The tree structure is also preserved.

Clearly, if Eq.~\eqref{eq:tensor_factor_graph} comes from a factor graph $P(\mathbf{X})$, any marginal distribution $m [x_i] = \sum_{\mathbf{x} \backslash x_i} \sum_{\mathbf{r}} T[\mathbf{x}, \mathbf{r}]$ corresponds exactly to a marginal $m (X_i)$ of $P (\mathbf{X})$.
Similar considerations hold if Eq.~\eqref{eq:tensor_factor_graph} defines a tensor network as in Eq.~\eqref{eq:general_tn}.
This observations motivate us to study inference on tensor factor graphs.

Given the exact correspondence with belief propagation we know that Eq.~\eqref{eq:fixed_point_tnbp} is exact on trees.
In complete analogy to the classical BP case, we can express marginals in terms of fixed point environments $\{ E^*_{s \rightarrow n} \}$ as
\begin{align}
  \label{eq:local_marginals_tensor_network_x}
  &m [x_i] = \frac{1}{Z_{\partial i} (E^*_{\partial i})}\sum_{r_{\partial i \backslash i_a}} T_i [x_i, r_{\partial i}] \bigotimes_{b \in \partial i \backslash a} E^*_{b \rightarrow i}[r_{b_i}], \\
  \label{eq:local_marginals_tensor_network_r}
  &m [r_{s_n}] = \frac{1}{Z_{sn} (E^*_{s\rightarrow n}, E^*_{n \rightarrow s})} E^*_{s\rightarrow n} [r_{s_n}] E^*_{n \rightarrow s} [r_{n_s}],
\end{align}
where $r_{s_n}$ and $r_{n_s}$ are the same index. $Z_{\partial i} (E^*_{\partial i})$ and $Z_{sn} (E^*_{s\rightarrow n}, E^*_{n \rightarrow s})$
are ``local partition functions'' that enforce the conventional L1 normalization of the marginal distributions $\sum_{x_i} m [x_i] = \sum_{r_{s_n}} m [r_{s_n}] = 1$.
Importantly, Eqs.~(\ref{eq:local_marginals_tensor_network_x},\ref{eq:local_marginals_tensor_network_r}) give an explicit expression for the marginals in terms of fixed-point environments.
A similar expression holds for local conditional distributions.
In Fig.~\ref{fig:tn_marginals}a and \ref{fig:tn_marginals}b, we provide the graphical representation of $m[x_i]$ and $m[r_{s_n}]$, where we omitted the normalization factors.
Graphically, the normalizations $Z_{\partial i} (E^*_{\partial i})$ and $Z_{sn} (E^*_{s\rightarrow n}, E^*_{n \rightarrow s})$ can be simply interpreted as a summation over all values of the open indices.


It is easy to see that the joint distribution can be written as a product of local marginals as
\begin{align}
  \label{eq:joint_dist_envs}
  T [\mathbf{x}, \mathbf{r}] \propto \prod_{i} m_i [x_i, r_{\partial i}] \prod_{j} m_j [r_j] ^{-1} 
\end{align}
where $m_j [r_j]$ was given in Eq.~\eqref{eq:local_marginals_tensor_network_r}, and the inverse is meant element-wise. The marginals $m_i [x_i, r _{\partial i}]$ are defined as
\begin{equation}
  \label{eq:marginal_region_envs}
  m_i [x_i, r _{\partial i}] =  \frac{1}{Z_{\partial i} (E^*_{\partial i})} T_i [x_i, r_{\partial i}] \bigotimes_{j \in \partial i } E^*_{j \rightarrow i} [r_{b_i}].
\end{equation}
In Fig.~\ref{fig:tn_marginals}c, we provide the graphical representation of $m_i [x_i, r _{\partial i}]$, up to the normalization factor $Z_{\partial i} (E^*_{\partial i})$.

To see how Eq.~\eqref{eq:joint_dist_envs} holds, we just need to insert Eqs.~(\ref{eq:marginal_region_envs},\ref{eq:local_marginals_tensor_network_r}) and then evaluate the joint tensor network on a specific value $T [\mathbf{x}^*, \mathbf{r}^*]$.
The tensor products appearing in the final expression transform into regular products and the environments appearing in $m_i [x_i, r _{\partial i}]$ and $m_j [r_j]$ cancel each other.
It is important to stress that Eq.~\eqref{eq:joint_dist_envs} is basis dependent.

By using Eq.~\eqref{eq:joint_dist_envs} we can express all physical quantities such as the internal energy, entropy, and free entropy (energy) in terms of local marginals and thus in terms of environments. For example, as it was already shown in Ref.~\cite{Alkabetz2021}, the Bethe free entropy takes the following form,
\begin{equation}\label{eq:tensor_bethe_free_entropy}
  \begin{split}
    \mathbf{F}_T (m) = &- \sum_{i} \sum_{x_i, r_{\partial i}} m_i [x_i, r _{\partial i}] \log \frac{m_i [x_i, r _{\partial i}]} {T_i [x_i, r_ {\partial i}]} \\
  &+ \sum_{j} \sum_{r_{\partial i}} m_j [r_j] \log m_j [r_j]. 
  \end{split}
\end{equation}
By using Eqs.~(\ref{eq:marginal_region_envs},\ref{eq:local_marginals_tensor_network_r}), we can once again express the marginals in terms of environments.
We obtain that the total free entropy can be written in terms of ``local partition functions'' as 
\begin{equation}\label{eq:tensor_bethe_free_entropy_envs}
  \begin{split}
    \mathbf{F}_T (E) &= \sum_{i \in V} \log Z_{\partial i} (E_{\partial i}) \\
    &- \sum_{j \in V_r} \log Z_{ij} (E_{i\rightarrow j}, E_{j \rightarrow i}). \\
  \end{split}
\end{equation}
The free entropies in Eqs.~(\ref{eq:tensor_bethe_free_entropy}, \ref{eq:tensor_bethe_free_entropy_envs}) can be seen as functions that, given a collection of tensors, output a scalar.
If the tensors correspond to the exact marginals or to the fixed-point environments, the scalar is exactly the free entropy of the original tensor factor graph.

The intimate connection between tensor factor graphs and regular factor graphs, and the explicit form of the free entropy above, allow us to trivially extend many fundamental results from the classical message-passing literature.
In particular, it is possible to extend the theorem by Yedidia et al.~\cite{Yedidia2001} which establishes a one-to-one correspondence between stationary points of the free entropy and fixed points of belief propagation.
While Eq.~\eqref{eq:tensor_bethe_free_entropy_envs} is exact only on trees, the result in Ref.~\cite{Yedidia2001} holds for arbitrary loopy graphs.
By employing a simple and elegant argument in terms of Lagrange multipliers, the authors showed that messages (environments) are fixed points of belief propagation if and only if they are zero gradient points of a Bethe free entropy similar to the one in Eq.~\eqref{eq:tensor_bethe_free_entropy_envs}.
Quite interestingly, as in the case of belief propagation, this correspondence defines a variational principle for the enviroments over the free entropy (energy) expressed as a sum of local terms.
This approach could provide alternative routes to tensor network contractions which exploit variational optimization of the associated free entropy. 

\section{One step Replica symmetry breaking}\label{sec:1rsb}

In the previous section we focused on tensor factor graphs supported on trees.
In this section we aim to relax that assumption, whilst assuming that the graph is sparse.

If a factor graph is supported on a tree, we know it admits a unique (stable) BP fixed point.
Therefore, the free entropy is concave.
If the graph has loops, this is no longer the case and, as the number of loops increases, there will be more and more BP fixed points~\footnote{The reader familiar with tensor networks, may a appreciate the connection with the general procedure of dealing with tensor networks with few loops. Namely, by cutting a given bond and storing the result as a sum over the corresponding singular values. Notice that, the sum grows exponentially with the numbers of cut bonds.}.
If the factor graph is not supported on a tree, we can still attempt to run BP in order to find some approximate fixed points.
In practice, this procedure works rather well for sparse graphs with loops.
However, as the graph becomes denser, BP ceases to converge and we need to resort to more sophisticated methods.

When we deal with loopy graphs, it is a rather common procedure to add correlations between environments or messages in order to construct better approximations for the marginal distributions.
The one-step Replica Symmetry Breaking (1RSB) cavity method~\cite{mezard_bethe_2001, mezard_cavity_2003} is a heuristic method of this kind where correlations between messages are modeled with a Boltzmann distribution.
Despite being heuristically (and some times rigorously) well motivated, it is not the only viable option.
Several inference algorithms on factor graphs and heuristic tensor-network contractions adopt different kind of strategies when it comes to model those correlations~\cite{ran2020tensor, Yedidia2001}.
The goal of this section is to translate the 1RSB cavity method from factor graphs to tensor networks.
In the literature, there exist several formulations of the 1RSB cavity method, here we follow Chapter 19 in Ref.~\cite{Mezard_Montanari_book_2009}.
We start by reviewing the method in the classical factor graph language, and then we translate it onto the framework introduced in the previous sections.

In what follows, we will call ``Bethe measure'' a distribution whose local marginals can be written in terms of approximate BP fixed points.
Thus, independently of the underlying graph, a Bethe measure admits an approximate quasi-solution of the BP equations.
And the marginal distribution of the Bethe measure can be approximated by the usual BP expressions, similar to what we derived in the previous section.
A precise definition of Bethe measure is beyond the scope of this work, further details can be found in Refs.~\cite{Mezard_Montanari_book_2009, talagrand2010mean} and references therein.

The 1RSB cavity method is based on the following three assumptions~\cite{Mezard_Montanari_book_2009, mezard_cavity_2003}:
\begin{enumerate}
\item There exist exponentially many quasi-solutions of the BP equations in the number of variables.
\item The canonical measure $P$ can be written as a convex combination of Bethe measures as
  \begin{equation}
    \label{eq:bethe_decomposition}
    P(\mathbf{X}) = \sum_{n} \omega_n \mu^n (\mathbf{X})
  \end{equation}
  where the weights $\omega_n = \exp(F_n)$ and $F_n$ is the free entropy of the corresponding Bethe measure $\mu^n$.
\item To the leading order, the number of Bethe measures equals the number of quasi solutions of BP. 
\end{enumerate}
The decomposition in Eq.~\eqref{eq:bethe_decomposition} can be understood as the pure state decomposition of $P(\mathbf{X})$.
Such interpretation is rather common in statistical mechanics.
For example, the thermal distribution of the ferromagnetic Ising model at low temperatures can be decomposed into the symmetric sum of two pure states: the ones with positive and the ones with negative expected magnetization.
More generally, any system with a global discrete symmetry would fall into this category where we have a finite number of pure states, independent of the number of variables.
The scenario where the number of pure states is finite and independent of system size is usually dubbed as ``replica symmetric'' (RS).
In the RS phase, the distribution itself is a Bethe measure, therefore local algorithms like BP converge efficiently to a fixed point.
The statistical mechanics community has identified other scenarios where the replica symmetric assumption does not hold.
In these cases, the probability distribution decomposes into a convex combination of exponentially many Bethe measures as in Eq.~\eqref{eq:bethe_decomposition}, and it is usually referred as ``one-step replica symmetry breaking'' (1RSB).
As the proliferation of fixed points may cause long and strong correlations between variables, BP will have a hard time to converge to a fixed point in the 1RSB phase.
The 1RSB cavity method aims to overcome this problem by studying the distribution over Bethe measures induced by the $\omega_n$'s in Eq.~\eqref{eq:bethe_decomposition}, rather than the original factor graph.

In order to quantitatively tackle the problem, let us rewrite Eq.~\eqref{eq:BP_first} in fixed-point form as $\nu_{ja}=f_i(\{\hat\nu_{bj}\}_{b \in \partial j \backslash a})$, and Eq.~\eqref{eq:BP_second} as $\hat\nu_{ai}=f_a(\{\nu_{ja}\}_{j \in \partial a \backslash i})$.
If the distribution admits many BP fixed points, we define as $\{\nu^n, \hat{\nu}^n \}$ the set of messages for the $n$-th fixed point.
To each fixed point we can associate a free entropy $F_n = \sum_a F_a(\{ \nu^n_{ja} \}) + \sum_i F_i (\{ \hat\nu^n_{aj} \}) - \sum_{ia} F_{ia}(\{ \nu^n_{ja}, \hat\nu^n_{aj} \})$, where $F_a$, $F_i$ and $F_{ia}$ take a simple analytical form, similar to Eq.~\eqref{eq:tensor_bethe_free_entropy}.

Once we assume we have exponentially many pure states, the prescription of the 1RSB cavity method is to assign a Boltzmann factor $\omega_n = \exp (F_n (\{\nu^n, \hat{\nu}^n \}))$ to each $\omega_n$ in Eq.~\eqref{eq:bethe_decomposition}.
The distribution induced by the $\omega_n$ on the fixed-point messages can be formally written as
\begin{equation}\label{eq:1RSB_factor_graph}
  \Psi(\nu, \hat\nu) = \mathcal{D} (\{ \nu, \hat\nu \}) \exp (\beta_P F (\{\nu, \hat{\nu} \})),
\end{equation}
where $\beta_P$ plays the role of an effective inverse temperature, and it is sometimes dubbed ``Parisi 1RSB parameter''.
The first term $\mathcal{D} (\{ \nu, \hat\nu \})$ in Eq.~\eqref{eq:1RSB_factor_graph} enforces that all messages satisfy the BP fixed-point equations.
Namely,
\begin{equation}\label{eq:BP_fp_constraint}
  \begin{split}
    \mathcal{D} (\{ \nu, \hat\nu \}) = &\prod_a \mathbb{I} \left(\hat\nu_{ai} = f_a(\{\nu_{ja}\}_{j \in \partial a \backslash i} \right) \times \\
  &\prod_i \mathbb{I} \left(\nu_{ja} = f_i(\{\hat\nu_{bj}\}_{b \in \partial j \backslash a}) \right),
  \end{split}
\end{equation}
where $\mathbb{I} (\cdot)$ is the indicator function.
The second term in Eq.~\eqref{eq:1RSB_factor_graph} can be also split into a product of terms that depend only on neighboring (local) messages as $\exp(\beta_P F)=\prod_a\exp(\beta_P F_a)\prod_i\exp(\beta_P F_i)\prod_{ia}\exp(\beta_P F_{ia})$, where we used the explicit decomposition of a free entropy for a fixed point. 
Clearly, the distribution $\Psi(\nu, \hat\nu)$ over fixed-point messages is itself a factor graph.
And it is easy to see that its graph retains the exact same topology of the original graph.
Eq.~\eqref{eq:1RSB_factor_graph} is often referred to as auxiliary factor graph.
The central assumption behind the 1RSB cavity method is that $\Psi(\nu, \hat\nu)$ can be better approximated by belief propagation than the original $P(\mathbf{X})$.
Namely, the auxiliary factor graph is ``more mean field'' than the original factor graph.

Since the auxiliary factor graph is still a factor graph, one can heuristically study it with the BP Eqs.~(\ref{eq:BP_first},\ref{eq:BP_second}) and hope they converge.
It has been observed that for certain problems this is indeed the case.
Then, the corresponding fixed points can be used to compute quantities related to the original factor graph.


In order to translate these concepts into the tensor network language, we assume that our tensor factor graph can be decomposed as a sum of exponentially many pure states.
However, we do not really need to assume that.
Any tensor network supported on a loopy graph can be decomposed into a linear combination of tensor networks supported on trees.
Assuming the graph is completely connected, one just needs to perform a singular value decomposition (SVD) on at least $\Sigma = N_e - N_v - 1$ bonds, where $N_v$ is the number of vertices and $N_e$ is the number of edges.
Because a connected graph is a tree if and only if it has $N_v - 1$ edges, we obtain a linear decomposition of our tensor network into a sum of tree tensor networks.
The number of trees can be computed analytically once the number of non zero singular values is known for each removed bond.
Let us call that number $\gamma_i$ for the $i$-th bond.
The total number of tree tensor networks is then equal to $\prod_{i=0}^{\Sigma} \gamma_i$ which is $O(\exp N)$ if $\Sigma$ is $O(N)$. 
Because each tree tensor network is supported on a tree, it is a Bethe measure and it admits an unique {\it exact} BP fixed point. 

The second assumption of the 1RSB cavity method states that the weight of each Bethe measure is equal to a Boltzmann weight.
This assumption is harder to justify in the tensor network language because the tree decomposition is not unique.
Furthermore, one could perform different decompositions, other than SVD, that may result to similar linear combinations.
It would be interesting to explore whether there exists decompositions that lead to the exact 1RSB Boltzmann weights.

Our goal is now to derive the tensor-network version of $\Psi (\nu, \hat{\nu})$ in Eq.~\eqref{eq:1RSB_factor_graph}.
Notice that the only information we need is one analytic expression for the fixed-point environments and one for the free entropy of that fixed point.
These expressions were both given in Sec.~\eqref{sec:tensor_factor_graph} in Eq.~\eqref{eq:fixed_point_tnbp} and Eq.~\eqref{eq:tensor_bethe_free_entropy_envs}, respectively.
Let us first simplify the notation and express Eq.~\eqref{eq:fixed_point_tnbp} as $E_{sn} = f_s(\{E_{ks} \}_{k \in \partial s \backslash n})$.
We can then proceed by defining the following auxiliary distribution over the environments
\begin{align}
  \label{eq:1rsb_tensor_network}
  T_{1RSB} (E) \cong \mathcal{D} (\{ E \}) e^{\beta_P \mathbf{F}_n (E)},
\end{align}
where $\mathcal{D} (\{ E \})$ can be again decomposed as a product of local terms
\begin{equation}\label{eq:TNBP_fp_constraint}
  \begin{split}
    \mathcal{D} (\{ E \}) = \prod_s \mathbb{I} \left(E_{sn} = f_s(\{E_{ks} \}_{k \in \partial s \backslash n}) \right).
  \end{split}
\end{equation}
The second factor in Eq.~\eqref{eq:1rsb_tensor_network} can be also split as a product of terms
\begin{equation}
  \label{eq:Boltzmann_factor_auxiliary_factor_graph}
  e^ {\beta_P \mathbf{F}_n (E)} = \prod_{i} \left( \frac{Z_{ij} (E_{i\rightarrow j}, E_{j \rightarrow i})}{Z_{\partial i} (E_{\partial i})}\right)^{\beta_P},
\end{equation}
where we used the explicit form of $\mathbf{F}_n$ from Eq.~\eqref{eq:tensor_bethe_free_entropy_envs}.

It is thus evident from Eqs.(\ref{eq:TNBP_fp_constraint},\ref{eq:Boltzmann_factor_auxiliary_factor_graph}) that the distribution over fixed-point environments in Eq.~\eqref{eq:1rsb_tensor_network} admits a {\it local} factorization with the same topology of the original tensor factor graph.

Following the previous discussion, we can use belief propagation or more sophisticated tensor contractions to study the auxiliary tensor networks.
It is important to stress that the distribution in Eq.~\eqref{eq:1rsb_tensor_network} could be defined on continuous domains.
Indeed, in the most general case, the linear constraints defined by the tensor BP Eq.~\eqref{eq:fixed_point_tnbp} clearly admit a continuous set of solutions.
This is a serious complication that should be tackled on a case-by-case basis.

There exist simple cases, however, where the fixed-point conditions admit solutions only on a finite discrete alphabet.
Let us suppose that this is the case and the alphabet contains $d$ elements and that we found the corresponding fixed point environments $\hat{E}_{sn} = (E^0_{sn}, \ldots , E^d_{sn})$ that satisfy the tensor BP equation.
If this is the case, the auxiliary tensor network will represent a probability distribution of the environments over a $d^{2 * \mathcal{E}}-$dimensional simplex, where $\mathcal{E}$ is the number of bonds in the original graph, and the factor two accounts for the number of environments per bond.
By grouping together terms that depend on the same environments in Eq.~\eqref{eq:1rsb_tensor_network}, it is easy to see that $T_{1RSB} (E)$ can be formally rewritten as a product of functions that depend on a few environments, namely a factor graph.
Such factor graph has the same form as the one in Eq.~\eqref{eq:factor_graph_tn_section}: it is a product of functions, and each function is defined on a discrete domain.
We can therefore study it with belief propagation, or map it onto a tensor network and test different tensor contractions.
The classical spin-glass community has identified a number of problems where the auxiliary model can be simplified this way.
In Sec.~\ref{sec:ksat}, we will deal with one of such examples.

\subsection{Higher orders of Replica Symmetry Breaking}\label{sec:higher_RSB}

In some cases, the assumption that the auxiliary model is a Bethe measure is not correct.
If, for example, the original graph is very dense, it is likely that the first step of replica symmetry breaking will not suffice, and simple mean-field algorithms like belief propagation will not converge.
Indeed, it could be the case that the auxiliary model itself breaks into a linear combination of many Bethe measures.
In such cases, one option is to explore higher orders of replica symmetry breaking.

As we outlined above, the auxiliary model can be interpreted as a conventional factor graph or tensor network.
Therefore, the recipe that produces a 1RSB auxiliary model from a factor graph can be conceptually bootstrapped to obtain the two-step replica symmetry breaking (2RSB) model.
The main technical bottleneck is hidden within the nature of the 1RSB step.
As it was commented above, the 1RSB model is a distribution over messages (or environments).
In the worst case scenario, such distribution is defined over a space of continuous probability distributions: the 1RSB auxiliary model is a distribution of distributions (recall that, each BP fixed point defines a distribution, and the 1RSB step defines a distribution over fixed points).
By bootstrapping the recipe above, one may incur a rather complicated mathematical object that is a distribution over a distribution of distributions.
Despite some attempts to study higher order RSB models~\cite{mezard1987spin}, only limited results have been produced.
It would be interesting to explore if the algebraic interpretation provided by tensor networks could give additional insights into this challenging problem.

There are two observations that make us hopeful.
The first is that, thanks to the multilinear representation of the problem, we can associate a finite and discrete Hilbert space to each environment.
Such space is spanned by a small set of basis vectors, and can be always characterized rather easily.
The second is the eigenvalue decomposition of the fixed-point Eq.~\eqref{eq:fixed_point_tnbp}.
There are many exact numerical methods that can be employed to solve such tensor eigenvalue problem.
Many of which are extensively used to study tensor networks~\cite{ran2020tensor}.
These methods can be employed to efficiently construct and characterize all local {\it eigen}-solutions of the fixed-point conditions, and facilitate the {\it numerical} construction of the 1RSB auxiliary model.

\section{An example:  {\it k}-SAT}\label{sec:ksat}

\begin{figure*}
    \includegraphics[width=\textwidth]{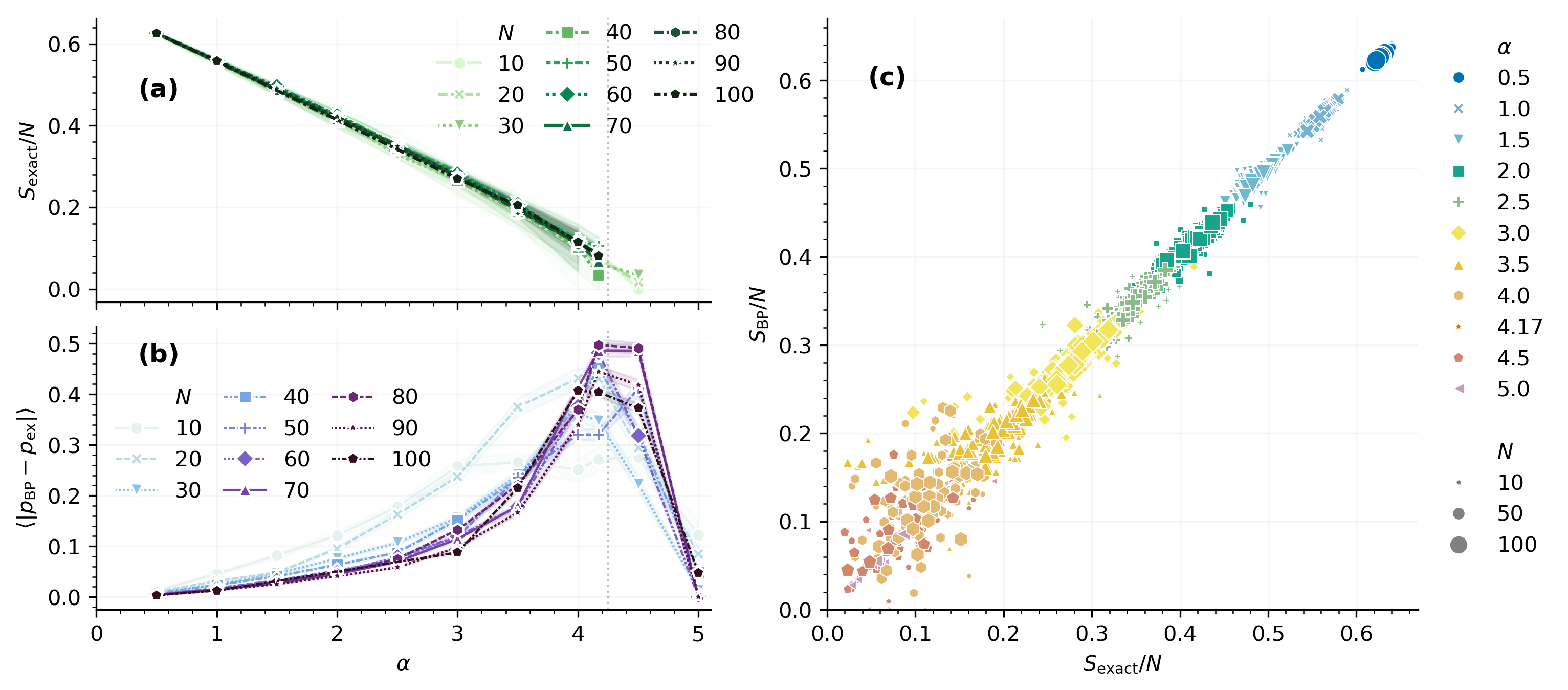}
    \caption{
    \textbf{
    The breakdown of belief propagation on $k$-SAT.~\cite{gitrepo, gray2018quimb}
    }
    \textbf{(a)}: average behavior of the exact free entropy per variable, $S/N$, for 3-SAT as a function of clause density, $\alpha$, and problem size, $N$. The approximate location of the SAT-UNSAT transition is marked.
    \textbf{(b)}: average distance between exact and BP-estimated variable marginals, as a function of $\alpha$ and $N$. For the purpose of this plot we consider BP to have found the exact marginal if it correctly identifies a problem as UNSAT.
    \textbf{(c)}: direct comparison of the exact and BP-estimated  free entropy per variable, plotted for varying $\alpha$ and $N$.
    All plots show statistics across 20 instances at each setting.
    }\label{fig:bp-vs-exact}
\end{figure*}

The 1RSB cavity method found extensive applications outside physics.
One of the first and most successful examples being combinatorial optimization.
Ideas from the theory of spin glasses and, in particular, one-step replica symmetry breaking are the backbone of the so-called Survey Propagation algorithm, the best deterministic {\it k}-SAT solver.
The interesting feature of Survey Propagation is that, not only does it produce heuristic to find SAT assignments, but it does so by exploiting a refined and detailed statistical description of the problem.
In this section we will introduce a tensor-network version of survey propagation.

The problem of finding a SAT assignment of a {\it k}-SAT instance can be formulated in terms of inference on a factor graph.
We can define an energy function $E_a(X_{\partial a}) = 1$ if $X_{\partial a}$ violates the clause $a$, and $E_a(X_{\partial a}) = 0$ otherwise.
Each variable can take two distinct boolean values, namely $|\mathcal{X}| = 2$ everywhere.
Whereas each clause $a$ is represented by a function $E_a(X_{\partial a})$, where $|\partial a| = K$.
For each clause $a$, there is only one configuration of $X_{\partial a}$ that violates the constraint.
Thus, if any of the variables $X_i$ with $i \in \partial a$ satisfies the clause $a$, then the clause is automatically satisfied. 
The problem of finding a SAT assignment is equivalent to finding a configuration $X^*$ such that
\begin{equation}
  \label{eq:energy_kSAT}
  E (X^*) = \sum_{a=1}^M E_a (X^*_{\partial a}) = 0.
\end{equation}
In statistical mechanics, one would model the Boltzmann distribution of the problem as
\begin{equation}
  \label{eq:thermal_ksat}
  \frac{e^{-\beta E(X)}}{Z}  = \frac{1}{Z} \prod_{a=1}^M \exp(-\beta E_a(X_{\partial a})),
\end{equation}
which becomes the uniform distribution over all SAT assignments in the limit $\beta \rightarrow \infty$
\begin{equation}
  \label{eq:dist_sat_assigns}
  \begin{split}
    P_{\rm SAT} (X) &= \lim_{\beta \rightarrow \infty} \frac{1}{Z} \prod_a \exp(-\beta E_a(X_{\partial a})) \\
    &= \frac{1}{Z} \prod_{a=1}^M 1 - E_a (X_{\partial a}) = \frac{1}{Z} \prod_{a=1}^M p_a (X_{\partial a}).
  \end{split}
\end{equation}
Notice that the factor graph in Eq.~\eqref{eq:dist_sat_assigns} has exactly the same form as the one in Eq.~\eqref{eq:factor_graph_tn_section}, we can therefore apply our results from the previous section to the problem of satisfiability.

In order to find a SAT assignment, we can attempt to sample a configuration from $P_{\rm SAT} (X)$.
As we discuss in the previous sections, one possible approach is to run BP on $P_{\rm SAT}(X)$ and use fixed-point information to recursively fix all variables to some value~\cite{Pumphrey2001SOLVINGTS, Mezard_Montanari_book_2009}.
As we discussed in the previous sections, BP is guaranteed to converge on trees.
For loopy graphs, as long as they are locally tree like and the correlations decay fast enough, we can hope to find approximate solutions to the BP equations.
One can easily check if BP found a solution to the problem by computing the energy of the corresponding configuration.
If the ratio $\alpha = M/N$ between number of clauses $M$ and number of variables $N$ becomes large, BP will eventually cease to converge. 
It is a well known fact that, for small values of $\alpha$, the space of SAT assignments is concentrated in one single cluster~\cite{krzakala2007gibbs}.
In that scenario, given a SAT assignment, one can explore the space of solutions by flipping a small number of variables: solutions are close in Hamming distance.
As $\alpha$ increases, the space of solutions shatter into many clusters which are far away from each other~\cite{krzakala2007gibbs}.
Given a solution in one cluster, it takes $O(N)$ flips to find solutions in another cluster.
The shattering of the space of solutions is one of the reasons why BP stops converging when the density of the clauses is high.
Intuitively, when $\alpha$ is small, BP has one unique fixed point.
Namely, the graph is sparse and locally tree-like.
The correlations decay fast enough such that the neighboring messages/environments are effectively not correlated.
However, as $\alpha$ crosses a certain transition value, the number of BP fixed points increases abruptly and the algorithm ceases to converge.

In Fig.~\ref{fig:bp-vs-exact} we numerically illustrate the breakdown of belief propagation for 3-SAT, by comparing to exact results for both variable marginals and free entropy per variable for small problem sizes.
We use a combination of tensor network contraction~\cite{gray2021hyper} (for low clause density) and the exact model counter \texttt{ganak}~\cite{SRSM19} to compute the reference exact marginals and entropy.
Fig.~\ref{fig:bp-vs-exact}(a) shows the exact behavior of the entropy per variable, which decays to 0 as the clause density, $\alpha$, approaches the SAT-UNSAT transition.
In Fig.~\ref{fig:bp-vs-exact}(b) we compare the BP-estimated marginals, computed using Eq.~\eqref{eq:local_marginals_tensor_network_x}, to the exact marginals. The clear divergence as $\alpha \rightarrow \sim 4.2$ indicates that BP has either failed to converge or converged to incorrect local marginals.
Finally in Fig.~\ref{fig:bp-vs-exact}(c) we directly compare the total entropy per variable, $S/N = \log Z / N$, which serves as a proxy for the ability of BP to estimate the full contracted value of the corresponding tensor network.
We see good correspondence for $\alpha < 3.5$, especially for larger $N$, but again this breaks down. 

In a series of groundbreaking results~\cite{mezard_cavity_2003, mezard_bethe_2001}, statistical physicists came up with new algorithms that exploit the solution space decomposition and greatly outperform simpler local algorithms like belief propagation.
These algorithms adopt the ideas discussed in the previous sections.

Survey Propagation (SP), exploits the proliferation of fixed points by assuming that the distribution of solutions decomposes into a linear combination of exponentially many clusters (or pure states), similar to Eq.~\eqref{eq:bethe_decomposition}.
SP was developed as the 1RSB message-passing algorithm corresponding to warning propagation (WP). The WP update rules in fixed point form are
\begin{align}
  \label{eq:WP_1}
  \nu_{i \rightarrow a} (X_i) &= \min \{ 1, \sum_{b \in \partial \backslash a } \hat{\nu}_{b \rightarrow i} (X_i) + C_{i \rightarrow a} \},\\
  \label{eq:WP_2}
    \hat{\nu}_{a \rightarrow i} (X_i) &= \min_{X_{\partial a \backslash i}} \{ E_a (X_{\partial a}) + \sum_{j \in \partial a \backslash i} \nu_{j \rightarrow a} (X_j) \} + \hat{C}_{a \rightarrow i}
\end{align}
where $C_{i \rightarrow a}$ and  $\hat{C}_{a \rightarrow i}$ are normalization constants.
The WP messages carry warnings about values that the corresponding variable should (or should not) take.
Eqs.~(\ref{eq:WP_1}, \ref{eq:WP_2}) admit fixed-point solutions in a discrete space where the messages take only binary values $\hat{\nu}_{b \rightarrow i} (X_i) \in \{ 0, 1\}$ and $\nu_{j \rightarrow a} (X_j) \in \{ 0, 1 \}$.
In this situation the 1RSB equations simplify considerably.
It has been empirically shown that SP can vastly outperform BP and WP as it can find SAT assignments efficiently also in the shattered phase, very close to the theoretical SAT-UNSAT transition~\cite{braunstein2005survey}.
Details about the connections between WP, BP and SP can be found in Refs.~\cite{braunstein2005survey, Mezard_Montanari_book_2009}.
Here we will just report the SP equations and map them onto tensor contractions.

In order to introduce the SP equations, let us define $\mathcal{C}_j^a = a \cup \mathcal{S}_{ja} \cup \mathcal{U}_{ja}$ as the set of all clauses that contain the variable $x_j$.
$\mathcal{S}_{ja}$ is the subset of all clauses that force $x_j$ in the same direction of $a$, and $\mathcal{U}_{ja}$ in the opposite direction.
Additionally, let us define four real and positive numbers ($\hat{Q}_{ia}$, $Q^U_{ia}$, $Q^S_{ia}$, $Q^*_{ia}$) that will play the role of the messages in survey propagation. $Q^S_{ia}$ is the probability that $x_i$ is forced by the clauses $b \in \partial i \backslash a$ to satisfy $a$,
$Q^U_{ia}$ is the probability that $x_i$ is forced by the clauses $b \in \partial i \backslash a$ to violate $a$,
$Q^*_{ia}$ is the probability that $x_i$ is not forced by the clauses $b \in \partial i \backslash a$,
and $\hat{Q}_{ai}$ is the probability that $x_i$ is forced by the clause $a$ to satisfy it.

The fixed-point SP equations can be written in the following way
\begin{align}
  \label{eq:sp_clause}
  \hat{Q}_{ai} &= \prod_{j \in \partial a \backslash i} Q^U_{ja},\\
  \label{eq:sp_U}
  Q^U_{ja} &\cong \mathcal{P}_{\mathcal{S}_{ja}} (\hat{Q}_{bj}) (1- \mathcal{P}_{\mathcal{U}_{ja}} (\hat{Q}_{bj})),\\
  \label{eq:sp_S}
  Q^S_{ja} &\cong \mathcal{P}_{\mathcal{U}_{ja}} (\hat{Q}_{bj}) (1- \mathcal{P}_{\mathcal{S}_{ja}} (\hat{Q}_{bj})),\\
  \label{eq:sp_star}
  Q^*_{ja} &\cong \mathcal{P}_{\mathcal{S}_{ja}} (\hat{Q}_{bj}) \mathcal{P}_{\mathcal{U}_{ja}} (\hat{Q}_{bj}),
\end{align}
where $Q^U_{ja} + Q^S_{ja} + Q^*_{ja} = 1$, and $\mathcal{P}_{\mathcal{S}_{ja}} (\hat{Q}_{bj}) = \prod_{b \in \mathcal{S}_{ja}} (1-\hat{Q}_{bj})$ and $\mathcal{P}_{\mathcal{U}_{ja}}(\hat{Q}_{bj}) = \prod_{b \in \mathcal{U}_{ja}} (1-\hat{Q}_{bj})$.
In the remainder, we will show that Eqs.~(\ref{eq:sp_clause}-\ref{eq:sp_star}) can be interpreted as regular tensorized BP equations on a simple tensor network.
The central observation is that the SP equations are multilinear in the input variables $Q_{ja}^U$ and $\hat{Q}_{bj}$.
Therefore, it is possible to rephrase them in terms of tensor contractions.
Our goal is to represent Eqs.~(\ref{eq:sp_clause}-\ref{eq:sp_star}) as a single fixed-point tensor contraction as Eq.~\eqref{eq:fixed_point_tnbp}.
First, let us group all SP messages into two types of environment vectors
\begin{equation}
  \label{eq:sp_env}
  \begin{split}
    E^{SP}_{ai} &= \begin{pmatrix} \hat{Q}_{ai} & 1-\hat{Q}_{ai} & 0 & 0 & 0 \end{pmatrix}, \\
  E^{SP}_{ja} &= \begin{pmatrix} 0 & 0 & Q^U_{ja} & Q^S_{ja} & Q^*_{ja} \end{pmatrix},
  \end{split}
\end{equation}
where $E^{SP}_{ai}$ is a clause-to-variable environment and $E^{SP}_{ja}$ is a variable-to-clause environment. 
The quantities $\mathcal{P}_{\mathcal{S}_{ja}} (\hat{Q}_{bj})$ and $\mathcal{P}_{\mathcal{U}_{ja}} (\hat{Q}_{bj})$ can be easily cast into a tensor contraction as
\begin{equation}
  \label{eq:tensor_PU_PS}
  \begin{split}
    \mathcal{P}_{\mathcal{S}_{ja}} (\hat{Q}_{bj}) &= \sum_{r_{\partial j \in \mathcal{S}_{ja}}} \mathcal{A}^{\mathcal{S}_{ja}}_{ja} [r_{\partial b}] \bigotimes E^{SP}_{bj} [r_{b_j}],\\
    \mathcal{P}_{\mathcal{U}_{ja}} (\hat{Q}_{bj}) &= \sum_{r_{\partial j \in \mathcal{U}_{ja}}} \mathcal{A}^{\mathcal{U}_{ja}}_{ja} [r_{\partial b}] \bigotimes E^{SP}_{bj} [r_{b_j}],
  \end{split}
\end{equation}
where we introduced the tensors $\mathcal{A}^{\mathcal{S}_{ja}}_{ja}$ and $\mathcal{A}^{\mathcal{U}_{ja}}_{ja}$ that are zero everywhere except for the single entry $\mathcal{A}[2, \ldots , 2]~=~1$, and have $|\mathcal{S}_{ja}|$ and $|\mathcal{U}_{ja}|$ indices, respectively.
Similarly, the first SP Eq.~\eqref{eq:sp_clause} can be written as a tensor contraction as
\begin{equation}
  \label{eq:tensor_SP_clause}
  \hat{Q}_{ai} = \sum_{r_{\partial a}} \mathcal{A}^{\mathcal{C}}_{aj} [r_{\partial a}] \bigotimes E^{SP}_{ia} [r_{i_a}],
\end{equation}
where $\mathcal{A}^{\mathcal{C}}_{aj}$ has $|\partial a| - 1$ indices and it is zero everywhere with exception of the one entry $\mathcal{A}^{\mathcal{C}}_{aj}[3,3,\ldots , 3] = 1$.

We can finally rewrite Eqs.~(\ref{eq:sp_clause}-\ref{eq:sp_star}) as a single tensor contraction as
\begin{equation}
  \label{eq:tensor_SP}
  E^{SP}_{sn} [r_{s_n}] = \sum_{r_{\partial s \backslash n_s}} \mathcal{A}_{sn} [r_{\partial s}] \bigotimes E^{SP}_{ks} [r_{s_k}],
\end{equation}
where $\mathcal{A}_{sn}$ is either a variable $\mathcal{A}_{ja}$ or a clause $\mathcal{A}_{aj}$ tensor defined as
\begin{align}
  \label{eq:local_tensor_SP_clause}
  \mathcal{A}_{a \rightarrow j} = ~& \mathbf{e}^1_j \otimes \mathcal{A}^{\mathcal{C}}_{aj} + \mathbf{e}^2_j \otimes (\mathbf{1}^{\mathcal{C}} - \mathcal{A}^{\mathcal{C}}_{aj})\\
  \label{eq:local_tensor_SP_variable}
  \mathcal{A}_{j \rightarrow a} = ~&\mathbf{e}^3_a \otimes \mathcal{A}_{ja}^{\mathcal{S}_{ja}} \otimes (\mathbf{1}^{\mathcal{U}_{ja}} - \mathcal{A}_{ja}^{\mathcal{U}_{ja}}) + \\
  &\mathbf{e}^4_a \otimes \mathcal{A}_{ja}^{\mathcal{U}_{ja}} \otimes (\mathbf{1}^{\mathcal{S}_{ja}} - \mathcal{A}_{ja}^{\mathcal{S}_{ja}}) + \nonumber\\
  &\mathbf{e}^5_a \otimes \mathcal{A}_{ja}^{\mathcal{U}_{ja}} \otimes \mathcal{A}_{ja}^{\mathcal{S}_{ja}}.\nonumber
\end{align}
The canonical basis vectors $\mathbf{e}^{k}_j = (0, \ldots, 1_k, \ldots, 0)$ ensure that the output environment $E^{SP}_{sn} = \sum_i E^{SP}_{sn}[k]~\mathbf{e}^k_n$ is populated in the correct dimensions.
The tensors $\mathbf{1}^{\mathcal{C}}$, $\mathbf{1}^{\mathcal{U}_{ja}}$ and $\mathbf{1}^{\mathcal{S}_{ja}}$ can be understood as identity tensors in the relevant subspaces.
Namely,
\begin{equation}
  \label{eq:id_tensors}
  \begin{split}
    \mathbf{1}^{\mathcal{C}} &= \bigotimes_{j \in \mathcal{C}} (\mathbf{e}_1^{j} + \mathbf{e}_2^{j})\\
    \mathbf{1}^{\mathcal{U}_{ja}} &= \bigotimes_{j \in \mathcal{U}_{ja}} (\mathbf{e}_3^{j} + \mathbf{e}_4^{j} + \mathbf{e}_5^{j})\\
    \mathbf{1}^{\mathcal{S}_{ja}} &= \bigotimes_{j \in \mathcal{S}_{ja}} (\mathbf{e}_3^{j} + \mathbf{e}_4^{j} + \mathbf{e}_5^{j}).\\
  \end{split}
\end{equation}

Given a {\it k}-SAT instance, one can use the definitions of $\mathcal{U}_{ja}$ and $\mathcal{S}_{ja}$ above to construct all the tensors that compose the auxiliary tensor network representing survey propagation.
Notice that all tensors involved are extremely sparse.
By exploiting the sparsity of the problem, one obtains exactly the original SP Eqs.~(\ref{eq:sp_clause}-\ref{eq:sp_star}).

\subsection{Decimation}

We can attempt to find a solution of a {\it k}-SAT instance by employing BP and SP.
In practice, one initializes all the environments (or messages) at random and runs the message passing equations until convergence.
If the algorithm converges to a fixed point, we can use the corresponding environments to fix the value of a variable.
For each variable $X_i$, we can use the fixed point messages to compute a quantity $b_i \in [-1, 1]$ called {\it bias}.
We select the variable $X_k$ with the largest absolute $b_k$, and fix it according to the sign of $b_k$.
The biases $b_i = p^0_i - p^1_i$ can be expressed in terms of fixed-point environments pointing towards variable $X_i$.

In the case of BP, $p_i^0 = \mu (X_i = 0)$ and $p_i^1 = \mu(X_i = 1)$, where $\mu(X_i)$ is the (approximate) single variable marginal computed from the fixed point environments.

For SP, 
\begin{equation}
  \label{eq:sp_decimation}
  \begin{split}
    p_i^0 &= \prod_{b \in \partial_1 i} (1-\hat{Q}_{bj}) \left[1 - \prod_{b \in \partial_0 i} (1-\hat{Q}_{bj})\right]\\
  p_i^1 &= \prod_{b \in \partial_0 i} (1-\hat{Q}_{bj}) \left[1 - \prod_{b \in \partial_1 i} (1-\hat{Q}_{bj})\right],
  \end{split}
\end{equation}
where $\partial_0 i$ ($\partial_1 i$) is the subset of clauses where $X_i$ appears negated (not negated).

The details of the algorithm are provided in Appendix~\ref{appdx:algorithms}.
Once a variable has been fixed, we remove it from the instance, together with all clauses that that are automatically satisfied by that choice.
This procedure is called ``decimation'' and it is iterated until all variables have been fixed to some value.
In the case of SP, decimation is iterated until the largest bias reaches a certain threshold.
The resulting decimated instance can then be solved with local algorithms like WalkSat or BP.

\begin{figure}[h]
\includegraphics[width=0.49\textwidth]{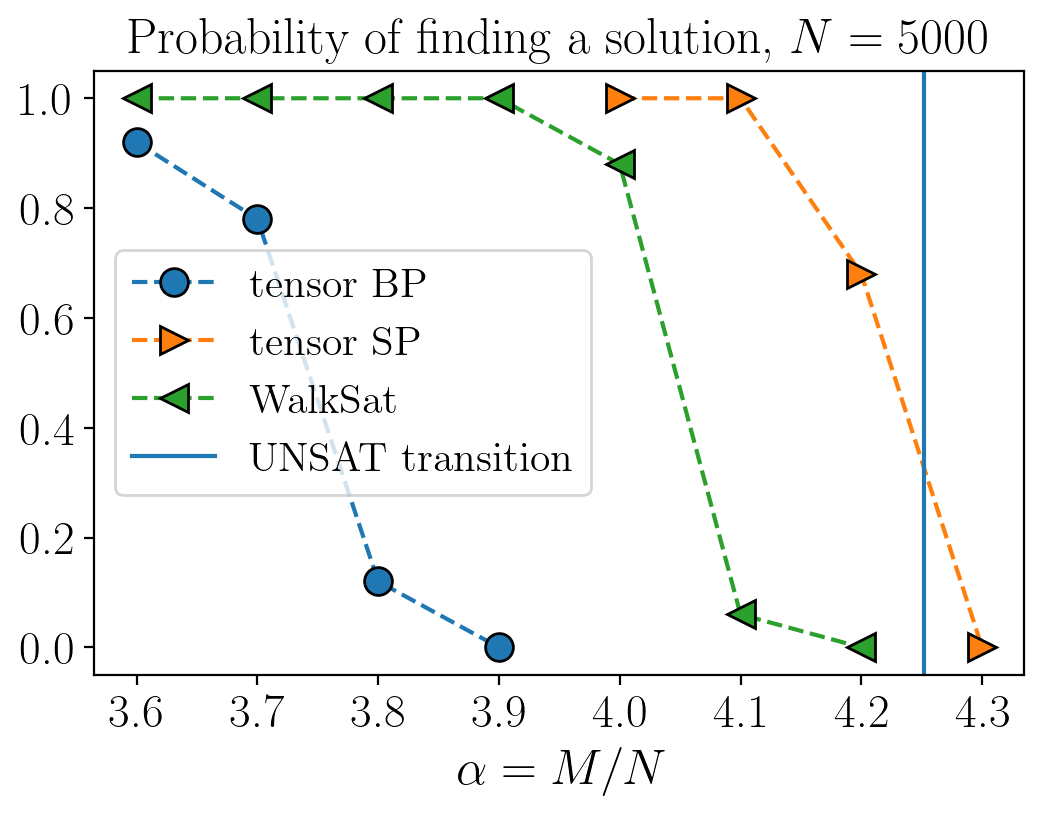}
\caption{{\bf Probability of finding a 3-SAT solution~\cite{gitrepo}}. We compute the probability to find a SAT assignment out of 50 random instances. BP and SP are initialized at random and iterated for at most 1000 iterations. Convergence is reached if all L1 distances between old and new environments are smaller than $10^{-3}$. We set the maximum number of iterations for WalkSat to $10^6$ and run it with several values for the flipping probabilities. SP is stopped once the largest bias is smaller than $10^{-3}$, then Walksat is called in the resulting decimated instance.}
\label{fig:solutions}
\end{figure}

In Fig.~\ref{fig:solutions} we compare the probability of finding a solution for 3-SAT random instances.
We fix the number of variables to $N = 5000$ and vary the density of clauses in proximity of the phase transition.
The figure shows the ratio of successful instances out of 50 random realizations
The exact algorithms are outlined in Appendix~\ref{appdx:algorithms}.

The data in Fig.~\ref{fig:solutions} has been collected on a single CPU without any sort of parallelization.
Notice that the graph associated to a {\it k}-SAT instance is definitely not regular and the number of distinct degrees can become very large, especially in the vicinity of the phase transition point.
Therefore, the strategy outlined in Sec.~\ref{sec:vtnbp} is not very efficient in this case.
In order to handle large instances, we used a sparse representation of the tensors.
It is important to stress that the results in Fig.~\ref{fig:solutions} are purely demonstrative and not at all surprising.
Since they rely on the exact mathematical mapping outlined above, these performances are completely expected and they have already been observed since more than two decades.
It has already been shown that SP can be easily pushed to much larger instances, up to $N \sim 10^7$ on single machines.
Because each local update affects only a few neighboring environments, these types of message-passing algorithms are very suitable for efficient parallelizations.

The interesting part of these results is the deep connection between the two theories.
There have been many attempts to efficiently solve combinatorial optimization problems with tensor networks.
Our approach, inspired from the classical theory of spin glasses, provides an alternative route that greatly outperforms previous proposals.
Both tensor networks and message passing algorithms have been extremely successful in their respective disciplines.
Our hope is that this connection may contribute to the two communities, and facilitate multidisciplinary adoption.

\section{Quantum tensor networks}\label{sec:quantum}

Tensor networks have been developed as a powerful tool to study quantum mechanical systems.
Researchers have extensively used them to represent quantum wavefunctions and operators in a way that is rather similar to the factor graphs discussed above.
A tensor-network representation of a quantum state is usually defined as 
\begin{equation}
  \label{eq:quantum_tn}
  \psi(x_1, \ldots, x_N) \propto \sum_{\{r\}} \prod_k T_k [x_k, r_{\partial k}],
\end{equation}
very similar to Eq.~\eqref{eq:tn_factor_graph}.
In this case, however, the tensors $T_k [x_k, r_{\partial k}]$ are not necessarily real and positive, their entries can be arbitrary complex numbers.
Therefore, the correct normalization of the state $\psi$ is the L2 norm, rather than the L1 norm as above.

In order to connect the two frameworks, consider the normalization constant $\| \psi(\mathbf{x}) \|^2 = \| \psi(x_1, \ldots, x_N) \|^2$ of the quantum state in Eq.~\eqref{eq:quantum_tn}.
This will play the same role of the partition function in the previous sections.
More explicitly,
\begin{equation}
  \label{eq:quantum_tn_x}
  \| \psi(\mathbf{x}) \|^2 = \sum_{\{ r, r' \}} \sum_{\{ x \}} \prod_k T_k [x_k, r_{\partial k}] \overline{T_k [x_k, r'_{\partial k}]},
\end{equation}
where the bar denotes the complex conjugation.
Since each variable $x_k$ appears only in one pair of local tensors, we can work out the corresponding summation to obtain
\begin{equation}
  \label{eq:quantum_norm}
  \| \psi(\mathbf{x}) \|^2 = \sum_{\{ r, r' \}} \prod_k W_k[r_{\partial k}, r'_{\partial k}],
\end{equation}
where $W_k[r_{\partial k}, r'_{\partial k}] = \sum_{x_k} T_k [x_k, r_{\partial k}] \overline{T_k [x_k, r'_{\partial k}]}$.
Notice that Eq.\eqref{eq:quantum_norm} has the exact same form as the tensor networks considered in the previous sections.
The only difference is that $W_k[r_{\partial k}, r'_{\partial k}]$ can take complex values.
The argument of the summation in Eq.~\eqref{eq:quantum_norm} can be therefore interpreted as a (complex) factor graph.
From the definition of the factors $W_k[r_{\partial k}, r'_{\partial k}]$, it is clear that each one of them must obey the symmetry $\overline{W_k[r_{\partial k}, r'_{\partial k}]} = W_k[r'_{\partial k}, r_{\partial k}]$.

In Ref.~\cite{Alkabetz2021}, the authors studied this type of tensor networks and they found a number of interesting connections between conventional contraction routines such as the simple update algorithm~\cite{jiang2008accurate} and belief propagation.
Remarkably, they proved the following theorem
\begin{theorem}[Alkabetz, Arad (2020)]
  \label{th:SU_BP}
Every trivial single update fixed point corresponds
to a belief propagation fixed point such that the local reduced density matrix, i.e. marginal, computed in both methods are identical.
\end{theorem}
The authors also showed that many of the quantities discussed in this paper, such as the marginals and the free entropy, can be extended to the quantum case.
For related work in the quantum regime, see also Refs.~\cite{sahu2022efficient, Cao2017, laumann2008cavity, leifer2008quantum}.

Similar to the classical case, let us consider the factor graph
\begin{equation}
  \label{eq:complex_factor_graph}
  \mathcal{W} [\mathbf{r}, \mathbf{r}'] = \prod_k W_k[r_{\partial k}, r'_{\partial k}],
\end{equation}
that is a product of positive semidefinite functions $W_k$, such that $\sum_{x, x'} h[x] W_k[x, x'] \overline{h[x']} \geq 0$ for any $h[x] > 0$.
The BP's fixed point conditions take a very simple form, similar to Eq.~\eqref{eq:fixed_point_tnbp}, as
\begin{equation}
  \label{eq:fixed_point_quantum_tnbp}
  E_{s \rightarrow n} [r_{s_n}, r'_{s_n}] = \sum_{r_{\partial s \backslash n_s}} W_s[r_{\partial s}, r'_{\partial s}]  \bigotimes_{k \in \partial s \backslash n} E_{k \rightarrow s} [r_{s_k}, r'_{s_k}].
\end{equation}
As a consequence, the environments can be chosen to satisfy $\overline{E_{s \rightarrow n} [r_{s_n}, r'_{s_n}]}~=~E_{s \rightarrow n} [r'_{s_n}, r_{s_n}]$.
These choices guaranteed that the resulting marginals, i.e. the reduced density matrices, are well defined.

The generalized class of factor graphs in Eq.~\eqref{eq:complex_factor_graph} emerges naturally from the theory of tensor networks and it is a special case of the double-edge factor graphs considered in Ref.~\cite{Cao2017}.
In order to connect the factor graph in Eq.~\eqref{eq:complex_factor_graph} with the results in the previous sections, we first need to introduce the relevant thermodynamic quantities such as internal energy, entropy, and free entropy.
We can follow the conventional statistical mechanics approach and define the partition function as 
\begin{equation}
  \label{eq:complex_partition_function}
  \mathcal{Z} = \| \psi(\mathbf{x}) \|^2 = \sum_{\{r, r'\}} \prod_k W_k[r_{\partial k}, r'_{\partial k}].
\end{equation}
Notice that Eq.~\eqref{eq:complex_partition_function} can be seen as an inner product $\mathbf{1}^T \mathcal{W} \mathbf{1}$, where $\mathbf{1} = (1, 1, \ldots, 1)$, and it is real and positive by construction.
Similar to the classical case, the internal energy and the entropy are defined as
\begin{align}
  \label{eq:complex_internal_energy}
  \mathcal{U} &= - \sum_{\{r, r'\}} \frac{\mathcal{W} [\mathbf{r}, \mathbf{r}']}{\mathcal{Z}} \sum_k \log W_k,\\
  \label{eq:complex_entropy}
  \mathcal{S} &= - \sum_{\{r, r'\}} \frac{\mathcal{W} [\mathbf{r}, \mathbf{r}']}{\mathcal{Z}} \sum_k \log \frac{W_k}{\mathcal{Z}},
\end{align}
where the complex logarithms are computed element wise.
As long as elements of $W_k$ do not lay on the logarithm's branch cut, the positivity of $W_k$ implies that $\mathcal{U}$ and $\mathcal{S}$ are real numbers.
It is not straightforward to show that $\mathcal{S} \geq 0$.
In fact it is easy to come up with pathological examples where $\mathcal{W} [\mathbf{r}, \mathbf{r}'] \neq 0$ and $\mathcal{Z} = 0$.
However, any realistic tensor network we considered has always exhibited a positive entropy.

By following the steps from the previous sections, it is easy to generalize many results to the complex case.
For instance, we can decompose $\mathcal{W}$ as a product of marginals as
\begin{equation}
  \label{eq:complex_factor_graph_marginals}
  \mathcal{W} [\mathbf{r}, \mathbf{r}'] \propto \prod_k M_{\partial k}[r_{\partial k}, r'_{\partial k}] \prod_{k'} M^{-1}_{k'} [r_{k'}, r_{k'}'],
\end{equation}
where
\begin{equation}\label{eq:quantum_marginals}
  \begin{split}
    &M_{\partial k}[r_{\partial k}, r'_{\partial k}] = \frac{1}{Z_{\partial k}} W_k [r_{\partial k}, r'_{\partial k}] \bigotimes_{k_s \in \partial k } E_{s \rightarrow k} [r_{k_s}, r'_{k_s }], \\
    &M_k[r_k, r_k'] = \frac{1}{Z_{sk}} E_{k \rightarrow s} [r_{k}, r'_{k}] E_{s \rightarrow k} [r_{k}, r'_{k}],\\
  \end{split}
\end{equation}
are expressed in terms of fixed point environments.
The local partition functions $Z_{\partial k}$ and $Z_{sk}$ enforce the conventional L2 normalization.

By using Eqs.~(\ref{eq:complex_internal_energy}, \ref{eq:complex_entropy}, \ref{eq:complex_factor_graph_marginals}), it is easy to show that the Bethe free entropy $\mathcal{F} = \mathcal{S} - \mathcal{U}$ of a tensor network can be written as
\begin{equation}\label{eq:quantum_tensor_bethe_free_entropy}
  \begin{split}
    \mathcal{F} (M) &= - \sum_{i} \sum_{x_i, r_{\partial i}} M_i [r _{\partial i}, r' _{\partial i}] \log \frac{M_i [r _{\partial i}, r' _{\partial i}]} {W_i [r_ {\partial i}, r'_ {\partial i}]}, \\
  &+ \sum_{j} \sum_{r_{\partial i}} M_j [r_j, r'_j] \log M_j [r_j, r'_j].
  \end{split}
\end{equation}
Notice that the free entropy of a quantum tensor network takes the exact same form of Eq.~\eqref{eq:tensor_bethe_free_entropy}~\cite{Alkabetz2021}.
And, as in the previous case, if we express the marginals in terms of environments we obtain
\begin{equation}\label{eq:quantum_tensor_bethe_free_entropy_envs}
  \begin{split}
    \mathcal{F} (E) &= \sum_{i} \log Z_{\partial i} (E_{\partial i}) \\
    &- \sum_{j} \log Z_{ij} (E_{i\rightarrow j}, E_{j \rightarrow i}).
  \end{split}
\end{equation}

The free entropy of a quantum tensor network is an interesting object per se.
Most of the results we surveyed above, extend naturally to the quantum case.
In particular, it has already been shown that the correspondence between stationary points of the free energy/entropy and BP fixed points~\cite{Yedidia2001} holds also in this case~\cite{Cao2017}.
That one-to-one correspondence provides an alternative route for tensor network contractions based on variational optimization of the free entropy (energy).
More generally, the thermodynamic quantities $\mathcal{U}$, $\mathcal{S}$, and $\mathcal{F}$ allow us to connect many of the familiar tools from statistical mechanics to the framework of tensor networks.
These quantities clearly carry information about the correlations and the macroscopic structure of the original quantum state.
It would be interesting to explore to which extent these quantities can be leveraged to study tensor networks and their contractions.

One of these examples is the 1RSB cavity method discussed above.
Indeed, Eq.~\eqref{eq:fixed_point_quantum_tnbp} and Eq.~\eqref{eq:quantum_tensor_bethe_free_entropy_envs} provide all necessary ingredients to construct the 1RSB auxiliary tensor network.
It is interesting to notice that, since the free entropy of a fixed point takes only real values, its 1RSB auxiliary distribution is ``classical'', it takes real and positive values only, despite the original tensor network can take complex values.

\subsection{Belief propagation, simple update and gauging}

Simple update~\cite{jiang2008accurate} is a common strategy to precondition a tensor network before contraction.
The procedure is sometimes referred to as {\it gauging}, since it is inspired from the canonical gauge fixing that is performed on tensor networks supported on tree graphs~\cite{fannes1992ground, friedman1997density, lepetit2000density, shi2006classical, tagliacozzo2009simulation, nagaj2008quantum, murg2010simulating, li2012efficient, nakatani2013efficient, pivzorn2013tree, gerster2014unconstrained, murg2015tree}.
In such case, we know that simple update and belief propagation have an unique fixed point, and that they coincide.
When the graph has loops, this is no longer the case, and simple update may converge to different fixed points depending on the initial conditions and the update strategy.
The correspondence between simple update and belief propagation in Theorem~\ref{th:SU_BP} allows us to draw some high level connections between the 1RSB cavity methods and typical contraction strategies.

Recall that the main intuition behind the 1RSB cavity method is to construct an auxiliary Boltzmann distribution over all BP fixed points.
Each fixed point is characterized by a set of self-consistent equations, and their weights are given by a Boltzmann factor of the corresponding free entropy.
In the language of tensor networks, the 1RSB auxiliary model can be interpreted as a distribution over all possible gauges, defined as fixed points of simple update.
Running simple update on the tensor network is equivalent to draw a random sample from such distribution.
Although reasonable, this approach does not exploit the structure of the full distribution and can easily result on biased samples.
An alternative approach, inspired from the RSB theory, is to explicitly construct and study the auxiliary model.
As it was shown in Sec.~\ref{sec:ksat}, sampling from the auxiliary model (SP in that case) could provide outstanding improvements.

\section{Discussion and future directions}

In this work we outlined a deep connection between the one-step replica symmetry breaking cavity method and tensor networks.
Our results can be summarized as follows.
({\it i}) We expanded and extended the results in Ref.~\cite{Alkabetz2021} to tensor networks coming from classical factor graphs. We showed how to express the free entropy of a tensor network locally in terms of fixed-point environments.
({\it ii}) We showed how the algorithmic theory of the 1RSB cavity method can be adapted to tensor networks and how to extend it to quantum-mechanical models.
({\it iii}) As a byproduct, we obtain heuristic tensor-network algorithms that can efficiently find solutions of large {\it k}-SAT instance in the vicinity of the SAT-UNSAT phase transition.

In the last decades there has been extensive research to develop sophisticated contraction strategies to study the quantum many-body problem.
Local algorithms like belief propagation or simple update are often used as good preconditioning for more advanced techniques that can exploit both the geometry of the graph and the local properties of the tensors.
The multilinear nature of tensor networks lends itself to a plethora of methods, adapted from linear algebra, that have been proven extremely effective.
We showed that these methods could be easily employed to test and improve the performance of conventional message-passing algorithms.
In particular, it would be interesting to use ideas discussed in this paper to develop heuristic methods capable of finding more frustrated sub-regions of the network and thus contract them with exact or approximate techniques that are known to outperform mean-field algorithms~\cite{wang2023tensor}.
Such regions could be for example identified by detecting the subgraphs where belief propagation has poorer performance, i.e. where the environments do not converge, or converge to inferior solutions.

At the same time, the classical theory of disordered systems is a very well-developed and established framework that has produced a number of interesting applications, such as the ones discussed in this paper.
Many of these ideas have not yet permeated into the quantum many-body theory or tensor networks.
Our results show that by leveraging these techniques, it is possible to develop tensor-network algorithms of practical interest, and potentially improve them.
On the other hand, they pave the way to extend the theory of tensor networks to large and sparse random graphs.

In the last forty years, the theory of Replica Symmetry Breaking (RSB) has produced an enormous amount of results in physics, mathematics and computer science.
Strengthening the connection with tensor networks, may enhance our understanding of quantum systems with many degrees of freedom and, ultimately, help us develop more advanced mean-field theories.

More recently, message-passing algorithms have experienced great development for a number of applications including compressed sensing, machine learning and optimization.
It would be interesting to explore to which extent the ideas developed in such scenarios can be combined with modern tensor contractions.

Finally, the solid connection between quantum tensor networks and statistical mechanics, outlined in Sec.~\ref{sec:quantum}, suggests that generalized thermodynamic concepts can be used to extract macroscopic information about quantum states as a function of microscopic quantities.
These ideas adapts very naturally to the case of complex factor graphs, and they allow the generalization of the conventional cavity method to the quantum regime.
The techniques discussed in this paper can provide additional tools to study novel quantum phases of matter in disordered materials.

\section{Acknowledgments}

We would like to acknowledge insightful discussions with Ada Altieri, Claudio Benzoni, Garnet K. Chan, Steve Flammia, Giacomo Giudice, Ashley Milsted, and Martin Schuetz. Work by J.G. (Quimb development and support)
was supported by the US National Science Foundation (NSF) via grant no. CHE-2102505.

\bibliography{library.bib}

\begin{thebibliography}{89}%
\makeatletter
\providecommand \@ifxundefined [1]{%
 \@ifx{#1\undefined}
}%
\providecommand \@ifnum [1]{%
 \ifnum #1\expandafter \@firstoftwo
 \else \expandafter \@secondoftwo
 \fi
}%
\providecommand \@ifx [1]{%
 \ifx #1\expandafter \@firstoftwo
 \else \expandafter \@secondoftwo
 \fi
}%
\providecommand \natexlab [1]{#1}%
\providecommand \enquote  [1]{``#1''}%
\providecommand \bibnamefont  [1]{#1}%
\providecommand \bibfnamefont [1]{#1}%
\providecommand \citenamefont [1]{#1}%
\providecommand \href@noop [0]{\@secondoftwo}%
\providecommand \href [0]{\begingroup \@sanitize@url \@href}%
\providecommand \@href[1]{\@@startlink{#1}\@@href}%
\providecommand \@@href[1]{\endgroup#1\@@endlink}%
\providecommand \@sanitize@url [0]{\catcode `\\12\catcode `\$12\catcode
  `\&12\catcode `\#12\catcode `\^12\catcode `\_12\catcode `\%12\relax}%
\providecommand \@@startlink[1]{}%
\providecommand \@@endlink[0]{}%
\providecommand \url  [0]{\begingroup\@sanitize@url \@url }%
\providecommand \@url [1]{\endgroup\@href {#1}{\urlprefix }}%
\providecommand \urlprefix  [0]{URL }%
\providecommand \Eprint [0]{\href }%
\providecommand \doibase [0]{http://dx.doi.org/}%
\providecommand \selectlanguage [0]{\@gobble}%
\providecommand \bibinfo  [0]{\@secondoftwo}%
\providecommand \bibfield  [0]{\@secondoftwo}%
\providecommand \translation [1]{[#1]}%
\providecommand \BibitemOpen [0]{}%
\providecommand \bibitemStop [0]{}%
\providecommand \bibitemNoStop [0]{.\EOS\space}%
\providecommand \EOS [0]{\spacefactor3000\relax}%
\providecommand \BibitemShut  [1]{\csname bibitem#1\endcsname}%
\let\auto@bib@innerbib\@empty
\bibitem [{\citenamefont {Cichocki}\ \emph {et~al.}(2016)\citenamefont
  {Cichocki}, \citenamefont {Lee}, \citenamefont {Oseledets}, \citenamefont
  {Phan}, \citenamefont {Zhao}, \citenamefont {Mandic} \emph
  {et~al.}}]{cichocki2016tensor}%
  \BibitemOpen
  \bibfield  {author} {\bibinfo {author} {\bibfnamefont {A.}~\bibnamefont
  {Cichocki}}, \bibinfo {author} {\bibfnamefont {N.}~\bibnamefont {Lee}},
  \bibinfo {author} {\bibfnamefont {I.}~\bibnamefont {Oseledets}}, \bibinfo
  {author} {\bibfnamefont {A.-H.}\ \bibnamefont {Phan}}, \bibinfo {author}
  {\bibfnamefont {Q.}~\bibnamefont {Zhao}}, \bibinfo {author} {\bibfnamefont
  {D.~P.}\ \bibnamefont {Mandic}},  \emph {et~al.},\ }\href@noop {} {\bibfield
  {journal} {\bibinfo  {journal} {Foundations and Trends{\textregistered} in
  Machine Learning}\ }\textbf {\bibinfo {volume} {9}},\ \bibinfo {pages} {249}
  (\bibinfo {year} {2016})}\BibitemShut {NoStop}%
\bibitem [{\citenamefont {Cichocki}\ \emph {et~al.}(2017)\citenamefont
  {Cichocki}, \citenamefont {Phan}, \citenamefont {Zhao}, \citenamefont {Lee},
  \citenamefont {Oseledets}, \citenamefont {Sugiyama}, \citenamefont {Mandic}
  \emph {et~al.}}]{cichocki2017tensor}%
  \BibitemOpen
  \bibfield  {author} {\bibinfo {author} {\bibfnamefont {A.}~\bibnamefont
  {Cichocki}}, \bibinfo {author} {\bibfnamefont {A.-H.}\ \bibnamefont {Phan}},
  \bibinfo {author} {\bibfnamefont {Q.}~\bibnamefont {Zhao}}, \bibinfo {author}
  {\bibfnamefont {N.}~\bibnamefont {Lee}}, \bibinfo {author} {\bibfnamefont
  {I.}~\bibnamefont {Oseledets}}, \bibinfo {author} {\bibfnamefont
  {M.}~\bibnamefont {Sugiyama}}, \bibinfo {author} {\bibfnamefont {D.~P.}\
  \bibnamefont {Mandic}},  \emph {et~al.},\ }\href@noop {} {\bibfield
  {journal} {\bibinfo  {journal} {Foundations and Trends{\textregistered} in
  Machine Learning}\ }\textbf {\bibinfo {volume} {9}},\ \bibinfo {pages} {431}
  (\bibinfo {year} {2017})}\BibitemShut {NoStop}%
\bibitem [{\citenamefont {Cirac}\ \emph {et~al.}(2021)\citenamefont {Cirac},
  \citenamefont {Perez-Garcia}, \citenamefont {Schuch},\ and\ \citenamefont
  {Verstraete}}]{cirac2021matrix}%
  \BibitemOpen
  \bibfield  {author} {\bibinfo {author} {\bibfnamefont {J.~I.}\ \bibnamefont
  {Cirac}}, \bibinfo {author} {\bibfnamefont {D.}~\bibnamefont {Perez-Garcia}},
  \bibinfo {author} {\bibfnamefont {N.}~\bibnamefont {Schuch}}, \ and\ \bibinfo
  {author} {\bibfnamefont {F.}~\bibnamefont {Verstraete}},\ }\href@noop {}
  {\bibfield  {journal} {\bibinfo  {journal} {Reviews of Modern Physics}\
  }\textbf {\bibinfo {volume} {93}},\ \bibinfo {pages} {045003} (\bibinfo
  {year} {2021})}\BibitemShut {NoStop}%
\bibitem [{\citenamefont {Or{\'u}s}(2014)}]{orus2014practical}%
  \BibitemOpen
  \bibfield  {author} {\bibinfo {author} {\bibfnamefont {R.}~\bibnamefont
  {Or{\'u}s}},\ }\href@noop {} {\bibfield  {journal} {\bibinfo  {journal}
  {Annals of physics}\ }\textbf {\bibinfo {volume} {349}},\ \bibinfo {pages}
  {117} (\bibinfo {year} {2014})}\BibitemShut {NoStop}%
\bibitem [{\citenamefont {Evenbly}(2022)}]{evenbly2022practical}%
  \BibitemOpen
  \bibfield  {author} {\bibinfo {author} {\bibfnamefont {G.}~\bibnamefont
  {Evenbly}},\ }\href@noop {} {\bibfield  {journal} {\bibinfo  {journal} {arXiv
  preprint arXiv:2202.02138}\ } (\bibinfo {year} {2022})}\BibitemShut {NoStop}%
\bibitem [{\citenamefont {Perez-Garcia}\ \emph {et~al.}(2006)\citenamefont
  {Perez-Garcia}, \citenamefont {Verstraete}, \citenamefont {Wolf},\ and\
  \citenamefont {Cirac}}]{perez2006matrix}%
  \BibitemOpen
  \bibfield  {author} {\bibinfo {author} {\bibfnamefont {D.}~\bibnamefont
  {Perez-Garcia}}, \bibinfo {author} {\bibfnamefont {F.}~\bibnamefont
  {Verstraete}}, \bibinfo {author} {\bibfnamefont {M.~M.}\ \bibnamefont
  {Wolf}}, \ and\ \bibinfo {author} {\bibfnamefont {J.~I.}\ \bibnamefont
  {Cirac}},\ }\href@noop {} {\bibfield  {journal} {\bibinfo  {journal} {arXiv
  preprint quant-ph/0608197}\ } (\bibinfo {year} {2006})}\BibitemShut {NoStop}%
\bibitem [{\citenamefont {Oseledets}(2011)}]{oseledets2011tensor}%
  \BibitemOpen
  \bibfield  {author} {\bibinfo {author} {\bibfnamefont {I.~V.}\ \bibnamefont
  {Oseledets}},\ }\href@noop {} {\bibfield  {journal} {\bibinfo  {journal}
  {SIAM Journal on Scientific Computing}\ }\textbf {\bibinfo {volume} {33}},\
  \bibinfo {pages} {2295} (\bibinfo {year} {2011})}\BibitemShut {NoStop}%
\bibitem [{\citenamefont {Verstraete}\ and\ \citenamefont
  {Cirac}(2004)}]{verstraete2004renormalization}%
  \BibitemOpen
  \bibfield  {author} {\bibinfo {author} {\bibfnamefont {F.}~\bibnamefont
  {Verstraete}}\ and\ \bibinfo {author} {\bibfnamefont {J.~I.}\ \bibnamefont
  {Cirac}},\ }\href@noop {} {\bibfield  {journal} {\bibinfo  {journal} {arXiv
  preprint cond-mat/0407066}\ } (\bibinfo {year} {2004})}\BibitemShut {NoStop}%
\bibitem [{\citenamefont {Vidal}(2007)}]{vidal2007entanglement}%
  \BibitemOpen
  \bibfield  {author} {\bibinfo {author} {\bibfnamefont {G.}~\bibnamefont
  {Vidal}},\ }\href@noop {} {\bibfield  {journal} {\bibinfo  {journal}
  {Physical review letters}\ }\textbf {\bibinfo {volume} {99}},\ \bibinfo
  {pages} {220405} (\bibinfo {year} {2007})}\BibitemShut {NoStop}%
\bibitem [{\citenamefont {Gray}\ and\ \citenamefont
  {Kourtis}(2021)}]{gray2021hyper}%
  \BibitemOpen
  \bibfield  {author} {\bibinfo {author} {\bibfnamefont {J.}~\bibnamefont
  {Gray}}\ and\ \bibinfo {author} {\bibfnamefont {S.}~\bibnamefont {Kourtis}},\
  }\href@noop {} {\bibfield  {journal} {\bibinfo  {journal} {Quantum}\ }\textbf
  {\bibinfo {volume} {5}},\ \bibinfo {pages} {410} (\bibinfo {year}
  {2021})}\BibitemShut {NoStop}%
\bibitem [{\citenamefont {Gray}\ and\ \citenamefont
  {Chan}(2022)}]{gray2022hyper}%
  \BibitemOpen
  \bibfield  {author} {\bibinfo {author} {\bibfnamefont {J.}~\bibnamefont
  {Gray}}\ and\ \bibinfo {author} {\bibfnamefont {G.~K.}\ \bibnamefont
  {Chan}},\ }\href@noop {} {\bibfield  {journal} {\bibinfo  {journal} {arXiv
  preprint arXiv:2206.07044}\ } (\bibinfo {year} {2022})}\BibitemShut {NoStop}%
\bibitem [{\citenamefont {M{\'e}zard}\ \emph {et~al.}(1987)\citenamefont
  {M{\'e}zard}, \citenamefont {Parisi},\ and\ \citenamefont
  {Virasoro}}]{mezard1987spin}%
  \BibitemOpen
  \bibfield  {author} {\bibinfo {author} {\bibfnamefont {M.}~\bibnamefont
  {M{\'e}zard}}, \bibinfo {author} {\bibfnamefont {G.}~\bibnamefont {Parisi}},
  \ and\ \bibinfo {author} {\bibfnamefont {M.~A.}\ \bibnamefont {Virasoro}},\
  }\href@noop {} {\emph {\bibinfo {title} {Spin glass theory and beyond: An
  Introduction to the Replica Method and Its Applications}}},\ Vol.~\bibinfo
  {volume} {9}\ (\bibinfo  {publisher} {World Scientific Publishing Company},\
  \bibinfo {year} {1987})\BibitemShut {NoStop}%
\bibitem [{\citenamefont {Nishimori}(2001)}]{nishimori2001statistical}%
  \BibitemOpen
  \bibfield  {author} {\bibinfo {author} {\bibfnamefont {H.}~\bibnamefont
  {Nishimori}},\ }\href@noop {} {\emph {\bibinfo {title} {Statistical physics
  of spin glasses and information processing: an introduction}}},\ \bibinfo
  {number} {111}\ (\bibinfo  {publisher} {Clarendon Press},\ \bibinfo {year}
  {2001})\BibitemShut {NoStop}%
\bibitem [{\citenamefont {Altieri}\ and\ \citenamefont
  {Baity-Jesi}(2023)}]{altieri2023introduction}%
  \BibitemOpen
  \bibfield  {author} {\bibinfo {author} {\bibfnamefont {A.}~\bibnamefont
  {Altieri}}\ and\ \bibinfo {author} {\bibfnamefont {M.}~\bibnamefont
  {Baity-Jesi}},\ }\href@noop {} {\bibfield  {journal} {\bibinfo  {journal}
  {arXiv preprint arXiv:2302.04842}\ } (\bibinfo {year} {2023})}\BibitemShut
  {NoStop}%
\bibitem [{\citenamefont {Talagrand}(2010)}]{talagrand2010mean}%
  \BibitemOpen
  \bibfield  {author} {\bibinfo {author} {\bibfnamefont {M.}~\bibnamefont
  {Talagrand}},\ }\href@noop {} {\emph {\bibinfo {title} {Mean field models for
  spin glasses}}},\ Vol.~\bibinfo {volume} {54}\ (\bibinfo  {publisher}
  {Springer Science \& Business Media},\ \bibinfo {year} {2010})\BibitemShut
  {NoStop}%
\bibitem [{\citenamefont {Mezard}\ and\ \citenamefont
  {Montanari}(2009)}]{Mezard_Montanari_book_2009}%
  \BibitemOpen
  \bibfield  {author} {\bibinfo {author} {\bibfnamefont {M.}~\bibnamefont
  {Mezard}}\ and\ \bibinfo {author} {\bibfnamefont {A.}~\bibnamefont
  {Montanari}},\ }\href@noop {} {\emph {\bibinfo {title} {Information, Physics,
  and Computation}}}\ (\bibinfo  {publisher} {Oxford University Press, Inc.},\
  \bibinfo {address} {USA},\ \bibinfo {year} {2009})\BibitemShut {NoStop}%
\bibitem [{\citenamefont {Zdeborov{\'a}}\ and\ \citenamefont
  {Krzakala}(2016)}]{zdeborova2016statistical}%
  \BibitemOpen
  \bibfield  {author} {\bibinfo {author} {\bibfnamefont {L.}~\bibnamefont
  {Zdeborov{\'a}}}\ and\ \bibinfo {author} {\bibfnamefont {F.}~\bibnamefont
  {Krzakala}},\ }\href@noop {} {\bibfield  {journal} {\bibinfo  {journal}
  {Advances in Physics}\ }\textbf {\bibinfo {volume} {65}},\ \bibinfo {pages}
  {453} (\bibinfo {year} {2016})}\BibitemShut {NoStop}%
\bibitem [{\citenamefont {M{\'e}zard}\ \emph {et~al.}(2002)\citenamefont
  {M{\'e}zard}, \citenamefont {Parisi},\ and\ \citenamefont
  {Zecchina}}]{mezard2002analytic}%
  \BibitemOpen
  \bibfield  {author} {\bibinfo {author} {\bibfnamefont {M.}~\bibnamefont
  {M{\'e}zard}}, \bibinfo {author} {\bibfnamefont {G.}~\bibnamefont {Parisi}},
  \ and\ \bibinfo {author} {\bibfnamefont {R.}~\bibnamefont {Zecchina}},\
  }\href@noop {} {\bibfield  {journal} {\bibinfo  {journal} {Science}\ }\textbf
  {\bibinfo {volume} {297}},\ \bibinfo {pages} {812} (\bibinfo {year}
  {2002})}\BibitemShut {NoStop}%
\bibitem [{\citenamefont {Montanari}(2021)}]{montanari2021optimization}%
  \BibitemOpen
  \bibfield  {author} {\bibinfo {author} {\bibfnamefont {A.}~\bibnamefont
  {Montanari}},\ }\href@noop {} {\bibfield  {journal} {\bibinfo  {journal}
  {SIAM Journal on Computing}\ ,\ \bibinfo {pages} {FOCS19}} (\bibinfo {year}
  {2021})}\BibitemShut {NoStop}%
\bibitem [{\citenamefont {Mezard}\ and\ \citenamefont
  {Parisi}(2001)}]{mezard_bethe_2001}%
  \BibitemOpen
  \bibfield  {author} {\bibinfo {author} {\bibfnamefont {M.}~\bibnamefont
  {Mezard}}\ and\ \bibinfo {author} {\bibfnamefont {G.}~\bibnamefont
  {Parisi}},\ }\href {\doibase 10.1007/PL00011099} {\bibfield  {journal}
  {\bibinfo  {journal} {The European Physical Journal B}\ }\textbf {\bibinfo
  {volume} {20}},\ \bibinfo {pages} {217} (\bibinfo {year} {2001})},\ \bibinfo
  {note} {arXiv: cond-mat/0009418}\BibitemShut {NoStop}%
\bibitem [{\citenamefont {Mézard}\ and\ \citenamefont
  {Parisi}(2003)}]{mezard_cavity_2003}%
  \BibitemOpen
  \bibfield  {author} {\bibinfo {author} {\bibfnamefont {M.}~\bibnamefont
  {Mézard}}\ and\ \bibinfo {author} {\bibfnamefont {G.}~\bibnamefont
  {Parisi}},\ }\href {\doibase 10.1023/A:1022221005097} {\bibfield  {journal}
  {\bibinfo  {journal} {Journal of Statistical Physics}\ }\textbf {\bibinfo
  {volume} {111}},\ \bibinfo {pages} {1} (\bibinfo {year} {2003})}\BibitemShut
  {NoStop}%
\bibitem [{\citenamefont {Robeva}\ and\ \citenamefont
  {Seigal}(2019)}]{robeva2019duality}%
  \BibitemOpen
  \bibfield  {author} {\bibinfo {author} {\bibfnamefont {E.}~\bibnamefont
  {Robeva}}\ and\ \bibinfo {author} {\bibfnamefont {A.}~\bibnamefont
  {Seigal}},\ }\href@noop {} {\bibfield  {journal} {\bibinfo  {journal}
  {Information and Inference: A Journal of the IMA}\ }\textbf {\bibinfo
  {volume} {8}},\ \bibinfo {pages} {273} (\bibinfo {year} {2019})}\BibitemShut
  {NoStop}%
\bibitem [{\citenamefont {Pearl}(1988)}]{pearl1988probabilistic}%
  \BibitemOpen
  \bibfield  {author} {\bibinfo {author} {\bibfnamefont {J.}~\bibnamefont
  {Pearl}},\ }\href@noop {} {\emph {\bibinfo {title} {Probabilistic reasoning
  in intelligent systems: networks of plausible inference}}}\ (\bibinfo
  {publisher} {Morgan kaufmann},\ \bibinfo {year} {1988})\BibitemShut {NoStop}%
\bibitem [{\citenamefont {Alkabetz}\ and\ \citenamefont
  {Arad}(2021)}]{Alkabetz2021}%
  \BibitemOpen
  \bibfield  {author} {\bibinfo {author} {\bibfnamefont {R.}~\bibnamefont
  {Alkabetz}}\ and\ \bibinfo {author} {\bibfnamefont {I.}~\bibnamefont
  {Arad}},\ }\href {\doibase 10.1103/PhysRevResearch.3.023073} {\bibfield
  {journal} {\bibinfo  {journal} {Phys. Rev. Research}\ }\textbf {\bibinfo
  {volume} {3}},\ \bibinfo {pages} {023073} (\bibinfo {year}
  {2021})}\BibitemShut {NoStop}%
\bibitem [{\citenamefont {M{\'e}zard}\ and\ \citenamefont
  {Zecchina}(2002)}]{mezard2002random}%
  \BibitemOpen
  \bibfield  {author} {\bibinfo {author} {\bibfnamefont {M.}~\bibnamefont
  {M{\'e}zard}}\ and\ \bibinfo {author} {\bibfnamefont {R.}~\bibnamefont
  {Zecchina}},\ }\href@noop {} {\bibfield  {journal} {\bibinfo  {journal}
  {Physical Review E}\ }\textbf {\bibinfo {volume} {66}},\ \bibinfo {pages}
  {056126} (\bibinfo {year} {2002})}\BibitemShut {NoStop}%
\bibitem [{\citenamefont {Braunstein}\ \emph {et~al.}(2005)\citenamefont
  {Braunstein}, \citenamefont {M{\'e}zard},\ and\ \citenamefont
  {Zecchina}}]{braunstein2005survey}%
  \BibitemOpen
  \bibfield  {author} {\bibinfo {author} {\bibfnamefont {A.}~\bibnamefont
  {Braunstein}}, \bibinfo {author} {\bibfnamefont {M.}~\bibnamefont
  {M{\'e}zard}}, \ and\ \bibinfo {author} {\bibfnamefont {R.}~\bibnamefont
  {Zecchina}},\ }\href@noop {} {\bibfield  {journal} {\bibinfo  {journal}
  {Random Structures \& Algorithms}\ }\textbf {\bibinfo {volume} {27}},\
  \bibinfo {pages} {201} (\bibinfo {year} {2005})}\BibitemShut {NoStop}%
\bibitem [{\citenamefont {Maneva}\ \emph {et~al.}(2007)\citenamefont {Maneva},
  \citenamefont {Mossel},\ and\ \citenamefont {Wainwright}}]{Maneva2007}%
  \BibitemOpen
  \bibfield  {author} {\bibinfo {author} {\bibfnamefont {E.}~\bibnamefont
  {Maneva}}, \bibinfo {author} {\bibfnamefont {E.}~\bibnamefont {Mossel}}, \
  and\ \bibinfo {author} {\bibfnamefont {M.~J.}\ \bibnamefont {Wainwright}},\
  }\href {\doibase 10.1145/1255443.1255445} {\bibfield  {journal} {\bibinfo
  {journal} {J. ACM}\ }\textbf {\bibinfo {volume} {54}},\ \bibinfo {pages}
  {17–es} (\bibinfo {year} {2007})}\BibitemShut {NoStop}%
\bibitem [{\citenamefont {Krzaka{\l}a}\ \emph {et~al.}(2007)\citenamefont
  {Krzaka{\l}a}, \citenamefont {Montanari}, \citenamefont {Ricci-Tersenghi},
  \citenamefont {Semerjian},\ and\ \citenamefont
  {Zdeborov{\'a}}}]{krzakala2007gibbs}%
  \BibitemOpen
  \bibfield  {author} {\bibinfo {author} {\bibfnamefont {F.}~\bibnamefont
  {Krzaka{\l}a}}, \bibinfo {author} {\bibfnamefont {A.}~\bibnamefont
  {Montanari}}, \bibinfo {author} {\bibfnamefont {F.}~\bibnamefont
  {Ricci-Tersenghi}}, \bibinfo {author} {\bibfnamefont {G.}~\bibnamefont
  {Semerjian}}, \ and\ \bibinfo {author} {\bibfnamefont {L.}~\bibnamefont
  {Zdeborov{\'a}}},\ }\href@noop {} {\bibfield  {journal} {\bibinfo  {journal}
  {Proceedings of the National Academy of Sciences}\ }\textbf {\bibinfo
  {volume} {104}},\ \bibinfo {pages} {10318} (\bibinfo {year}
  {2007})}\BibitemShut {NoStop}%
\bibitem [{\citenamefont {Marino}\ \emph {et~al.}(2016)\citenamefont {Marino},
  \citenamefont {Parisi},\ and\ \citenamefont
  {Ricci-Tersenghi}}]{marino2016backtracking}%
  \BibitemOpen
  \bibfield  {author} {\bibinfo {author} {\bibfnamefont {R.}~\bibnamefont
  {Marino}}, \bibinfo {author} {\bibfnamefont {G.}~\bibnamefont {Parisi}}, \
  and\ \bibinfo {author} {\bibfnamefont {F.}~\bibnamefont {Ricci-Tersenghi}},\
  }\href@noop {} {\bibfield  {journal} {\bibinfo  {journal} {Nature
  communications}\ }\textbf {\bibinfo {volume} {7}},\ \bibinfo {pages} {1}
  (\bibinfo {year} {2016})}\BibitemShut {NoStop}%
\bibitem [{\citenamefont {Barbier}\ \emph {et~al.}(2013)\citenamefont
  {Barbier}, \citenamefont {Krzakala}, \citenamefont {Zdeborov{\'a}},\ and\
  \citenamefont {Zhang}}]{barbier2013hard}%
  \BibitemOpen
  \bibfield  {author} {\bibinfo {author} {\bibfnamefont {J.}~\bibnamefont
  {Barbier}}, \bibinfo {author} {\bibfnamefont {F.}~\bibnamefont {Krzakala}},
  \bibinfo {author} {\bibfnamefont {L.}~\bibnamefont {Zdeborov{\'a}}}, \ and\
  \bibinfo {author} {\bibfnamefont {P.}~\bibnamefont {Zhang}},\ }in\ \href@noop
  {} {\emph {\bibinfo {booktitle} {Journal of Physics: Conference Series}}},\
  Vol.\ \bibinfo {volume} {473}\ (\bibinfo {organization} {IOP Publishing},\
  \bibinfo {year} {2013})\ p.\ \bibinfo {pages} {012021}\BibitemShut {NoStop}%
\bibitem [{\citenamefont {M{\'e}zard}\ and\ \citenamefont
  {Parisi}(1986)}]{mezard1986mean}%
  \BibitemOpen
  \bibfield  {author} {\bibinfo {author} {\bibfnamefont {M.}~\bibnamefont
  {M{\'e}zard}}\ and\ \bibinfo {author} {\bibfnamefont {G.}~\bibnamefont
  {Parisi}},\ }\href@noop {} {\bibfield  {journal} {\bibinfo  {journal} {EPL
  (Europhysics Letters)}\ }\textbf {\bibinfo {volume} {2}},\ \bibinfo {pages}
  {913} (\bibinfo {year} {1986})}\BibitemShut {NoStop}%
\bibitem [{\citenamefont {Mulet}\ \emph {et~al.}(2002)\citenamefont {Mulet},
  \citenamefont {Pagnani}, \citenamefont {Weigt},\ and\ \citenamefont
  {Zecchina}}]{mulet2002coloring}%
  \BibitemOpen
  \bibfield  {author} {\bibinfo {author} {\bibfnamefont {R.}~\bibnamefont
  {Mulet}}, \bibinfo {author} {\bibfnamefont {A.}~\bibnamefont {Pagnani}},
  \bibinfo {author} {\bibfnamefont {M.}~\bibnamefont {Weigt}}, \ and\ \bibinfo
  {author} {\bibfnamefont {R.}~\bibnamefont {Zecchina}},\ }\href@noop {}
  {\bibfield  {journal} {\bibinfo  {journal} {Physical review letters}\
  }\textbf {\bibinfo {volume} {89}},\ \bibinfo {pages} {268701} (\bibinfo
  {year} {2002})}\BibitemShut {NoStop}%
\bibitem [{\citenamefont {Braunstein}\ \emph {et~al.}(2003)\citenamefont
  {Braunstein}, \citenamefont {Mulet}, \citenamefont {Pagnani}, \citenamefont
  {Weigt},\ and\ \citenamefont {Zecchina}}]{braunstein2003polynomial}%
  \BibitemOpen
  \bibfield  {author} {\bibinfo {author} {\bibfnamefont {A.}~\bibnamefont
  {Braunstein}}, \bibinfo {author} {\bibfnamefont {R.}~\bibnamefont {Mulet}},
  \bibinfo {author} {\bibfnamefont {A.}~\bibnamefont {Pagnani}}, \bibinfo
  {author} {\bibfnamefont {M.}~\bibnamefont {Weigt}}, \ and\ \bibinfo {author}
  {\bibfnamefont {R.}~\bibnamefont {Zecchina}},\ }\href@noop {} {\bibfield
  {journal} {\bibinfo  {journal} {Physical Review E}\ }\textbf {\bibinfo
  {volume} {68}},\ \bibinfo {pages} {036702} (\bibinfo {year}
  {2003})}\BibitemShut {NoStop}%
\bibitem [{\citenamefont {Sourlas}(1989)}]{sourlas1989spin}%
  \BibitemOpen
  \bibfield  {author} {\bibinfo {author} {\bibfnamefont {N.}~\bibnamefont
  {Sourlas}},\ }\href@noop {} {\bibfield  {journal} {\bibinfo  {journal}
  {Nature}\ }\textbf {\bibinfo {volume} {339}},\ \bibinfo {pages} {693}
  (\bibinfo {year} {1989})}\BibitemShut {NoStop}%
\bibitem [{\citenamefont
  {Montanari}(2001{\natexlab{a}})}]{montanari2001finite}%
  \BibitemOpen
  \bibfield  {author} {\bibinfo {author} {\bibfnamefont {A.}~\bibnamefont
  {Montanari}},\ }in\ \href@noop {} {\emph {\bibinfo {booktitle} {PROCEEDINGS
  OF THE ANNUAL ALLERTON CONFERENCE ON COMMUNICATION CONTROL AND COMPUTING}}},\
  Vol.~\bibinfo {volume} {39}\ (\bibinfo {organization} {The University;
  1998},\ \bibinfo {year} {2001})\ pp.\ \bibinfo {pages} {655--661}\BibitemShut
  {NoStop}%
\bibitem [{\citenamefont
  {Montanari}(2001{\natexlab{b}})}]{montanari2001glassy}%
  \BibitemOpen
  \bibfield  {author} {\bibinfo {author} {\bibfnamefont {A.}~\bibnamefont
  {Montanari}},\ }\href@noop {} {\bibfield  {journal} {\bibinfo  {journal} {The
  European Physical Journal B-Condensed Matter and Complex Systems}\ }\textbf
  {\bibinfo {volume} {23}},\ \bibinfo {pages} {121} (\bibinfo {year}
  {2001}{\natexlab{b}})}\BibitemShut {NoStop}%
\bibitem [{\citenamefont {Metropolis}\ and\ \citenamefont
  {Ulam}(1949)}]{metropolis1949monte}%
  \BibitemOpen
  \bibfield  {author} {\bibinfo {author} {\bibfnamefont {N.}~\bibnamefont
  {Metropolis}}\ and\ \bibinfo {author} {\bibfnamefont {S.}~\bibnamefont
  {Ulam}},\ }\href@noop {} {\bibfield  {journal} {\bibinfo  {journal} {Journal
  of the American statistical association}\ }\textbf {\bibinfo {volume} {44}},\
  \bibinfo {pages} {335} (\bibinfo {year} {1949})}\BibitemShut {NoStop}%
\bibitem [{\citenamefont {Hastings}(1970)}]{hastings1970monte}%
  \BibitemOpen
  \bibfield  {author} {\bibinfo {author} {\bibfnamefont {W.~K.}\ \bibnamefont
  {Hastings}},\ }\href@noop {} {\  (\bibinfo {year} {1970})}\BibitemShut
  {NoStop}%
\bibitem [{\citenamefont {Kirkpatrick}\ \emph {et~al.}(1983)\citenamefont
  {Kirkpatrick}, \citenamefont {Gelatt~Jr},\ and\ \citenamefont
  {Vecchi}}]{kirkpatrick1983optimization}%
  \BibitemOpen
  \bibfield  {author} {\bibinfo {author} {\bibfnamefont {S.}~\bibnamefont
  {Kirkpatrick}}, \bibinfo {author} {\bibfnamefont {C.~D.}\ \bibnamefont
  {Gelatt~Jr}}, \ and\ \bibinfo {author} {\bibfnamefont {M.~P.}\ \bibnamefont
  {Vecchi}},\ }\href@noop {} {\bibfield  {journal} {\bibinfo  {journal}
  {science}\ }\textbf {\bibinfo {volume} {220}},\ \bibinfo {pages} {671}
  (\bibinfo {year} {1983})}\BibitemShut {NoStop}%
\bibitem [{\citenamefont {Earl}\ and\ \citenamefont
  {Deem}(2005)}]{earl2005parallel}%
  \BibitemOpen
  \bibfield  {author} {\bibinfo {author} {\bibfnamefont {D.~J.}\ \bibnamefont
  {Earl}}\ and\ \bibinfo {author} {\bibfnamefont {M.~W.}\ \bibnamefont
  {Deem}},\ }\href@noop {} {\bibfield  {journal} {\bibinfo  {journal} {Physical
  Chemistry Chemical Physics}\ }\textbf {\bibinfo {volume} {7}},\ \bibinfo
  {pages} {3910} (\bibinfo {year} {2005})}\BibitemShut {NoStop}%
\bibitem [{\citenamefont {Zhu}\ \emph {et~al.}(2015)\citenamefont {Zhu},
  \citenamefont {Ochoa},\ and\ \citenamefont {Katzgraber}}]{zhu2015efficient}%
  \BibitemOpen
  \bibfield  {author} {\bibinfo {author} {\bibfnamefont {Z.}~\bibnamefont
  {Zhu}}, \bibinfo {author} {\bibfnamefont {A.~J.}\ \bibnamefont {Ochoa}}, \
  and\ \bibinfo {author} {\bibfnamefont {H.~G.}\ \bibnamefont {Katzgraber}},\
  }\href@noop {} {\bibfield  {journal} {\bibinfo  {journal} {Physical review
  letters}\ }\textbf {\bibinfo {volume} {115}},\ \bibinfo {pages} {077201}
  (\bibinfo {year} {2015})}\BibitemShut {NoStop}%
\bibitem [{\citenamefont {Wang}\ \emph {et~al.}(2015)\citenamefont {Wang},
  \citenamefont {Machta},\ and\ \citenamefont
  {Katzgraber}}]{wang2015comparing}%
  \BibitemOpen
  \bibfield  {author} {\bibinfo {author} {\bibfnamefont {W.}~\bibnamefont
  {Wang}}, \bibinfo {author} {\bibfnamefont {J.}~\bibnamefont {Machta}}, \ and\
  \bibinfo {author} {\bibfnamefont {H.~G.}\ \bibnamefont {Katzgraber}},\
  }\href@noop {} {\bibfield  {journal} {\bibinfo  {journal} {Physical Review
  E}\ }\textbf {\bibinfo {volume} {92}},\ \bibinfo {pages} {013303} (\bibinfo
  {year} {2015})}\BibitemShut {NoStop}%
\bibitem [{\citenamefont {Barzegar}\ \emph {et~al.}(2018)\citenamefont
  {Barzegar}, \citenamefont {Pattison}, \citenamefont {Wang},\ and\
  \citenamefont {Katzgraber}}]{barzegar2018optimization}%
  \BibitemOpen
  \bibfield  {author} {\bibinfo {author} {\bibfnamefont {A.}~\bibnamefont
  {Barzegar}}, \bibinfo {author} {\bibfnamefont {C.}~\bibnamefont {Pattison}},
  \bibinfo {author} {\bibfnamefont {W.}~\bibnamefont {Wang}}, \ and\ \bibinfo
  {author} {\bibfnamefont {H.~G.}\ \bibnamefont {Katzgraber}},\ }\href@noop {}
  {\bibfield  {journal} {\bibinfo  {journal} {Physical Review E}\ }\textbf
  {\bibinfo {volume} {98}},\ \bibinfo {pages} {053308} (\bibinfo {year}
  {2018})}\BibitemShut {NoStop}%
\bibitem [{\citenamefont {Barzegar}\ \emph {et~al.}(2021)\citenamefont
  {Barzegar}, \citenamefont {Kankani}, \citenamefont {Mandr{\`a}},\ and\
  \citenamefont {Katzgraber}}]{barzegar2021optimization}%
  \BibitemOpen
  \bibfield  {author} {\bibinfo {author} {\bibfnamefont {A.}~\bibnamefont
  {Barzegar}}, \bibinfo {author} {\bibfnamefont {A.}~\bibnamefont {Kankani}},
  \bibinfo {author} {\bibfnamefont {S.}~\bibnamefont {Mandr{\`a}}}, \ and\
  \bibinfo {author} {\bibfnamefont {H.~G.}\ \bibnamefont {Katzgraber}},\
  }\href@noop {} {\bibfield  {journal} {\bibinfo  {journal} {Physical Review
  E}\ }\textbf {\bibinfo {volume} {104}},\ \bibinfo {pages} {035302} (\bibinfo
  {year} {2021})}\BibitemShut {NoStop}%
\bibitem [{\citenamefont {Mohseni}\ \emph {et~al.}(2021)\citenamefont
  {Mohseni}, \citenamefont {Eppens}, \citenamefont {Strumpfer}, \citenamefont
  {Marino}, \citenamefont {Denchev}, \citenamefont {Ho}, \citenamefont
  {Isakov}, \citenamefont {Boixo}, \citenamefont {Ricci-Tersenghi},\ and\
  \citenamefont {Neven}}]{mohseni2021nonequilibrium}%
  \BibitemOpen
  \bibfield  {author} {\bibinfo {author} {\bibfnamefont {M.}~\bibnamefont
  {Mohseni}}, \bibinfo {author} {\bibfnamefont {D.}~\bibnamefont {Eppens}},
  \bibinfo {author} {\bibfnamefont {J.}~\bibnamefont {Strumpfer}}, \bibinfo
  {author} {\bibfnamefont {R.}~\bibnamefont {Marino}}, \bibinfo {author}
  {\bibfnamefont {V.}~\bibnamefont {Denchev}}, \bibinfo {author} {\bibfnamefont
  {A.~K.}\ \bibnamefont {Ho}}, \bibinfo {author} {\bibfnamefont {S.~V.}\
  \bibnamefont {Isakov}}, \bibinfo {author} {\bibfnamefont {S.}~\bibnamefont
  {Boixo}}, \bibinfo {author} {\bibfnamefont {F.}~\bibnamefont
  {Ricci-Tersenghi}}, \ and\ \bibinfo {author} {\bibfnamefont {H.}~\bibnamefont
  {Neven}},\ }\href@noop {} {\bibfield  {journal} {\bibinfo  {journal} {arXiv
  preprint arXiv:2111.13628}\ } (\bibinfo {year} {2021})}\BibitemShut {NoStop}%
\bibitem [{\citenamefont {Schuetz}\ \emph
  {et~al.}(2022{\natexlab{a}})\citenamefont {Schuetz}, \citenamefont
  {Brubaker},\ and\ \citenamefont {Katzgraber}}]{schuetz2022combinatorial}%
  \BibitemOpen
  \bibfield  {author} {\bibinfo {author} {\bibfnamefont {M.~J.}\ \bibnamefont
  {Schuetz}}, \bibinfo {author} {\bibfnamefont {J.~K.}\ \bibnamefont
  {Brubaker}}, \ and\ \bibinfo {author} {\bibfnamefont {H.~G.}\ \bibnamefont
  {Katzgraber}},\ }\href@noop {} {\bibfield  {journal} {\bibinfo  {journal}
  {Nature Machine Intelligence}\ }\textbf {\bibinfo {volume} {4}},\ \bibinfo
  {pages} {367} (\bibinfo {year} {2022}{\natexlab{a}})}\BibitemShut {NoStop}%
\bibitem [{\citenamefont {Angelini}\ and\ \citenamefont
  {Ricci-Tersenghi}(2022)}]{angelini2022modern}%
  \BibitemOpen
  \bibfield  {author} {\bibinfo {author} {\bibfnamefont {M.~C.}\ \bibnamefont
  {Angelini}}\ and\ \bibinfo {author} {\bibfnamefont {F.}~\bibnamefont
  {Ricci-Tersenghi}},\ }\href@noop {} {\bibfield  {journal} {\bibinfo
  {journal} {Nature Machine Intelligence}\ ,\ \bibinfo {pages} {1}} (\bibinfo
  {year} {2022})}\BibitemShut {NoStop}%
\bibitem [{\citenamefont {Schuetz}\ \emph
  {et~al.}(2022{\natexlab{b}})\citenamefont {Schuetz}, \citenamefont
  {Brubaker},\ and\ \citenamefont {Katzgraber}}]{schuetz2022reply}%
  \BibitemOpen
  \bibfield  {author} {\bibinfo {author} {\bibfnamefont {M.~J.}\ \bibnamefont
  {Schuetz}}, \bibinfo {author} {\bibfnamefont {J.~K.}\ \bibnamefont
  {Brubaker}}, \ and\ \bibinfo {author} {\bibfnamefont {H.~G.}\ \bibnamefont
  {Katzgraber}},\ }\href@noop {} {\bibfield  {journal} {\bibinfo  {journal}
  {Nature Machine Intelligence}\ ,\ \bibinfo {pages} {1}} (\bibinfo {year}
  {2022}{\natexlab{b}})}\BibitemShut {NoStop}%
\bibitem [{\citenamefont {Schuetz}\ \emph
  {et~al.}(2022{\natexlab{c}})\citenamefont {Schuetz}, \citenamefont
  {Brubaker}, \citenamefont {Zhu},\ and\ \citenamefont
  {Katzgraber}}]{schuetz2022graph}%
  \BibitemOpen
  \bibfield  {author} {\bibinfo {author} {\bibfnamefont {M.~J.}\ \bibnamefont
  {Schuetz}}, \bibinfo {author} {\bibfnamefont {J.~K.}\ \bibnamefont
  {Brubaker}}, \bibinfo {author} {\bibfnamefont {Z.}~\bibnamefont {Zhu}}, \
  and\ \bibinfo {author} {\bibfnamefont {H.~G.}\ \bibnamefont {Katzgraber}},\
  }\href@noop {} {\bibfield  {journal} {\bibinfo  {journal} {Physical Review
  Research}\ }\textbf {\bibinfo {volume} {4}},\ \bibinfo {pages} {043131}
  (\bibinfo {year} {2022}{\natexlab{c}})}\BibitemShut {NoStop}%
\bibitem [{\citenamefont {Albash}\ and\ \citenamefont
  {Lidar}(2018)}]{albash2018adiabatic}%
  \BibitemOpen
  \bibfield  {author} {\bibinfo {author} {\bibfnamefont {T.}~\bibnamefont
  {Albash}}\ and\ \bibinfo {author} {\bibfnamefont {D.~A.}\ \bibnamefont
  {Lidar}},\ }\href@noop {} {\bibfield  {journal} {\bibinfo  {journal} {Reviews
  of Modern Physics}\ }\textbf {\bibinfo {volume} {90}},\ \bibinfo {pages}
  {015002} (\bibinfo {year} {2018})}\BibitemShut {NoStop}%
\bibitem [{\citenamefont {Farhi}\ \emph {et~al.}(2001)\citenamefont {Farhi},
  \citenamefont {Goldstone}, \citenamefont {Gutmann}, \citenamefont {Lapan},
  \citenamefont {Lundgren},\ and\ \citenamefont {Preda}}]{farhi2001quantum}%
  \BibitemOpen
  \bibfield  {author} {\bibinfo {author} {\bibfnamefont {E.}~\bibnamefont
  {Farhi}}, \bibinfo {author} {\bibfnamefont {J.}~\bibnamefont {Goldstone}},
  \bibinfo {author} {\bibfnamefont {S.}~\bibnamefont {Gutmann}}, \bibinfo
  {author} {\bibfnamefont {J.}~\bibnamefont {Lapan}}, \bibinfo {author}
  {\bibfnamefont {A.}~\bibnamefont {Lundgren}}, \ and\ \bibinfo {author}
  {\bibfnamefont {D.}~\bibnamefont {Preda}},\ }\href@noop {} {\bibfield
  {journal} {\bibinfo  {journal} {Science}\ }\textbf {\bibinfo {volume}
  {292}},\ \bibinfo {pages} {472} (\bibinfo {year} {2001})}\BibitemShut
  {NoStop}%
\bibitem [{\citenamefont {McClean}\ \emph {et~al.}(2021)\citenamefont
  {McClean}, \citenamefont {Harrigan}, \citenamefont {Mohseni}, \citenamefont
  {Rubin}, \citenamefont {Jiang}, \citenamefont {Boixo}, \citenamefont
  {Smelyanskiy}, \citenamefont {Babbush},\ and\ \citenamefont
  {Neven}}]{mcclean2021low}%
  \BibitemOpen
  \bibfield  {author} {\bibinfo {author} {\bibfnamefont {J.~R.}\ \bibnamefont
  {McClean}}, \bibinfo {author} {\bibfnamefont {M.~P.}\ \bibnamefont
  {Harrigan}}, \bibinfo {author} {\bibfnamefont {M.}~\bibnamefont {Mohseni}},
  \bibinfo {author} {\bibfnamefont {N.~C.}\ \bibnamefont {Rubin}}, \bibinfo
  {author} {\bibfnamefont {Z.}~\bibnamefont {Jiang}}, \bibinfo {author}
  {\bibfnamefont {S.}~\bibnamefont {Boixo}}, \bibinfo {author} {\bibfnamefont
  {V.~N.}\ \bibnamefont {Smelyanskiy}}, \bibinfo {author} {\bibfnamefont
  {R.}~\bibnamefont {Babbush}}, \ and\ \bibinfo {author} {\bibfnamefont
  {H.}~\bibnamefont {Neven}},\ }\href@noop {} {\bibfield  {journal} {\bibinfo
  {journal} {PRX Quantum}\ }\textbf {\bibinfo {volume} {2}},\ \bibinfo {pages}
  {030312} (\bibinfo {year} {2021})}\BibitemShut {NoStop}%
\bibitem [{\citenamefont {Mandra}\ \emph {et~al.}(2016)\citenamefont {Mandra},
  \citenamefont {Zhu}, \citenamefont {Wang}, \citenamefont {Perdomo-Ortiz},\
  and\ \citenamefont {Katzgraber}}]{mandra2016strengths}%
  \BibitemOpen
  \bibfield  {author} {\bibinfo {author} {\bibfnamefont {S.}~\bibnamefont
  {Mandra}}, \bibinfo {author} {\bibfnamefont {Z.}~\bibnamefont {Zhu}},
  \bibinfo {author} {\bibfnamefont {W.}~\bibnamefont {Wang}}, \bibinfo {author}
  {\bibfnamefont {A.}~\bibnamefont {Perdomo-Ortiz}}, \ and\ \bibinfo {author}
  {\bibfnamefont {H.~G.}\ \bibnamefont {Katzgraber}},\ }\href@noop {}
  {\bibfield  {journal} {\bibinfo  {journal} {Physical Review A}\ }\textbf
  {\bibinfo {volume} {94}},\ \bibinfo {pages} {022337} (\bibinfo {year}
  {2016})}\BibitemShut {NoStop}%
\bibitem [{\citenamefont {Karimi}\ \emph {et~al.}(2017)\citenamefont {Karimi},
  \citenamefont {Rosenberg},\ and\ \citenamefont
  {Katzgraber}}]{karimi2017effective}%
  \BibitemOpen
  \bibfield  {author} {\bibinfo {author} {\bibfnamefont {H.}~\bibnamefont
  {Karimi}}, \bibinfo {author} {\bibfnamefont {G.}~\bibnamefont {Rosenberg}}, \
  and\ \bibinfo {author} {\bibfnamefont {H.~G.}\ \bibnamefont {Katzgraber}},\
  }\href@noop {} {\bibfield  {journal} {\bibinfo  {journal} {Physical Review
  E}\ }\textbf {\bibinfo {volume} {96}},\ \bibinfo {pages} {043312} (\bibinfo
  {year} {2017})}\BibitemShut {NoStop}%
\bibitem [{\citenamefont {Mandra}\ and\ \citenamefont
  {Katzgraber}(2018)}]{mandra2018deceptive}%
  \BibitemOpen
  \bibfield  {author} {\bibinfo {author} {\bibfnamefont {S.}~\bibnamefont
  {Mandra}}\ and\ \bibinfo {author} {\bibfnamefont {H.~G.}\ \bibnamefont
  {Katzgraber}},\ }\href@noop {} {\bibfield  {journal} {\bibinfo  {journal}
  {Quantum Science and Technology}\ }\textbf {\bibinfo {volume} {3}},\ \bibinfo
  {pages} {04LT01} (\bibinfo {year} {2018})}\BibitemShut {NoStop}%
\bibitem [{\citenamefont {Aramon}\ \emph {et~al.}(2019)\citenamefont {Aramon},
  \citenamefont {Rosenberg}, \citenamefont {Valiante}, \citenamefont
  {Miyazawa}, \citenamefont {Tamura},\ and\ \citenamefont
  {Katzgraber}}]{aramon2019physics}%
  \BibitemOpen
  \bibfield  {author} {\bibinfo {author} {\bibfnamefont {M.}~\bibnamefont
  {Aramon}}, \bibinfo {author} {\bibfnamefont {G.}~\bibnamefont {Rosenberg}},
  \bibinfo {author} {\bibfnamefont {E.}~\bibnamefont {Valiante}}, \bibinfo
  {author} {\bibfnamefont {T.}~\bibnamefont {Miyazawa}}, \bibinfo {author}
  {\bibfnamefont {H.}~\bibnamefont {Tamura}}, \ and\ \bibinfo {author}
  {\bibfnamefont {H.~G.}\ \bibnamefont {Katzgraber}},\ }\href@noop {}
  {\bibfield  {journal} {\bibinfo  {journal} {Frontiers in Physics}\ }\textbf
  {\bibinfo {volume} {7}},\ \bibinfo {pages} {48} (\bibinfo {year}
  {2019})}\BibitemShut {NoStop}%
\bibitem [{\citenamefont {Pichler}\ \emph {et~al.}(2018)\citenamefont
  {Pichler}, \citenamefont {Wang}, \citenamefont {Zhou}, \citenamefont {Choi},\
  and\ \citenamefont {Lukin}}]{pichler2018quantum}%
  \BibitemOpen
  \bibfield  {author} {\bibinfo {author} {\bibfnamefont {H.}~\bibnamefont
  {Pichler}}, \bibinfo {author} {\bibfnamefont {S.-T.}\ \bibnamefont {Wang}},
  \bibinfo {author} {\bibfnamefont {L.}~\bibnamefont {Zhou}}, \bibinfo {author}
  {\bibfnamefont {S.}~\bibnamefont {Choi}}, \ and\ \bibinfo {author}
  {\bibfnamefont {M.~D.}\ \bibnamefont {Lukin}},\ }\href@noop {} {\bibfield
  {journal} {\bibinfo  {journal} {arXiv preprint arXiv:1808.10816}\ } (\bibinfo
  {year} {2018})}\BibitemShut {NoStop}%
\bibitem [{\citenamefont {Zhou}\ \emph {et~al.}(2020)\citenamefont {Zhou},
  \citenamefont {Wang}, \citenamefont {Choi}, \citenamefont {Pichler},\ and\
  \citenamefont {Lukin}}]{zhou2020quantum}%
  \BibitemOpen
  \bibfield  {author} {\bibinfo {author} {\bibfnamefont {L.}~\bibnamefont
  {Zhou}}, \bibinfo {author} {\bibfnamefont {S.-T.}\ \bibnamefont {Wang}},
  \bibinfo {author} {\bibfnamefont {S.}~\bibnamefont {Choi}}, \bibinfo {author}
  {\bibfnamefont {H.}~\bibnamefont {Pichler}}, \ and\ \bibinfo {author}
  {\bibfnamefont {M.~D.}\ \bibnamefont {Lukin}},\ }\href@noop {} {\bibfield
  {journal} {\bibinfo  {journal} {Physical Review X}\ }\textbf {\bibinfo
  {volume} {10}},\ \bibinfo {pages} {021067} (\bibinfo {year}
  {2020})}\BibitemShut {NoStop}%
\bibitem [{\citenamefont {Ebadi}\ \emph {et~al.}(2022)\citenamefont {Ebadi},
  \citenamefont {Keesling}, \citenamefont {Cain}, \citenamefont {Wang},
  \citenamefont {Levine}, \citenamefont {Bluvstein}, \citenamefont {Semeghini},
  \citenamefont {Omran}, \citenamefont {Liu}, \citenamefont {Samajdar} \emph
  {et~al.}}]{ebadi2022quantum}%
  \BibitemOpen
  \bibfield  {author} {\bibinfo {author} {\bibfnamefont {S.}~\bibnamefont
  {Ebadi}}, \bibinfo {author} {\bibfnamefont {A.}~\bibnamefont {Keesling}},
  \bibinfo {author} {\bibfnamefont {M.}~\bibnamefont {Cain}}, \bibinfo {author}
  {\bibfnamefont {T.~T.}\ \bibnamefont {Wang}}, \bibinfo {author}
  {\bibfnamefont {H.}~\bibnamefont {Levine}}, \bibinfo {author} {\bibfnamefont
  {D.}~\bibnamefont {Bluvstein}}, \bibinfo {author} {\bibfnamefont
  {G.}~\bibnamefont {Semeghini}}, \bibinfo {author} {\bibfnamefont
  {A.}~\bibnamefont {Omran}}, \bibinfo {author} {\bibfnamefont {J.-G.}\
  \bibnamefont {Liu}}, \bibinfo {author} {\bibfnamefont {R.}~\bibnamefont
  {Samajdar}},  \emph {et~al.},\ }\href@noop {} {\bibfield  {journal} {\bibinfo
   {journal} {Science}\ ,\ \bibinfo {pages} {eabo6587}} (\bibinfo {year}
  {2022})}\BibitemShut {NoStop}%
\bibitem [{\citenamefont {Kourtis}\ \emph {et~al.}(2019)\citenamefont
  {Kourtis}, \citenamefont {Chamon}, \citenamefont {Mucciolo},\ and\
  \citenamefont {Ruckenstein}}]{kourtis2019fast}%
  \BibitemOpen
  \bibfield  {author} {\bibinfo {author} {\bibfnamefont {S.}~\bibnamefont
  {Kourtis}}, \bibinfo {author} {\bibfnamefont {C.}~\bibnamefont {Chamon}},
  \bibinfo {author} {\bibfnamefont {E.}~\bibnamefont {Mucciolo}}, \ and\
  \bibinfo {author} {\bibfnamefont {A.}~\bibnamefont {Ruckenstein}},\
  }\href@noop {} {\bibfield  {journal} {\bibinfo  {journal} {SciPost Physics}\
  }\textbf {\bibinfo {volume} {7}},\ \bibinfo {pages} {060} (\bibinfo {year}
  {2019})}\BibitemShut {NoStop}%
\bibitem [{\citenamefont {Kissinger}\ and\ \citenamefont
  {Meichanetzidis}(2020)}]{kissinger2020tensor}%
  \BibitemOpen
  \bibfield  {author} {\bibinfo {author} {\bibfnamefont {A.}~\bibnamefont
  {Kissinger}}\ and\ \bibinfo {author} {\bibfnamefont {K.}~\bibnamefont
  {Meichanetzidis}},\ }\href@noop {} {\  (\bibinfo {year} {2020})}\BibitemShut
  {NoStop}%
\bibitem [{\citenamefont {Rams}\ \emph {et~al.}(2021)\citenamefont {Rams},
  \citenamefont {Mohseni}, \citenamefont {Eppens}, \citenamefont
  {Ja\l{}owiecki},\ and\ \citenamefont {Gardas}}]{Rams2021}%
  \BibitemOpen
  \bibfield  {author} {\bibinfo {author} {\bibfnamefont {M.~M.}\ \bibnamefont
  {Rams}}, \bibinfo {author} {\bibfnamefont {M.}~\bibnamefont {Mohseni}},
  \bibinfo {author} {\bibfnamefont {D.}~\bibnamefont {Eppens}}, \bibinfo
  {author} {\bibfnamefont {K.}~\bibnamefont {Ja\l{}owiecki}}, \ and\ \bibinfo
  {author} {\bibfnamefont {B.}~\bibnamefont {Gardas}},\ }\href {\doibase
  10.1103/PhysRevE.104.025308} {\bibfield  {journal} {\bibinfo  {journal}
  {Phys. Rev. E}\ }\textbf {\bibinfo {volume} {104}},\ \bibinfo {pages}
  {025308} (\bibinfo {year} {2021})}\BibitemShut {NoStop}%
\bibitem [{\citenamefont {Liu}\ \emph {et~al.}(2021)\citenamefont {Liu},
  \citenamefont {Wang},\ and\ \citenamefont {Zhang}}]{liu2021tropical}%
  \BibitemOpen
  \bibfield  {author} {\bibinfo {author} {\bibfnamefont {J.-G.}\ \bibnamefont
  {Liu}}, \bibinfo {author} {\bibfnamefont {L.}~\bibnamefont {Wang}}, \ and\
  \bibinfo {author} {\bibfnamefont {P.}~\bibnamefont {Zhang}},\ }\href@noop {}
  {\bibfield  {journal} {\bibinfo  {journal} {Physical Review Letters}\
  }\textbf {\bibinfo {volume} {126}},\ \bibinfo {pages} {090506} (\bibinfo
  {year} {2021})}\BibitemShut {NoStop}%
\bibitem [{\citenamefont {Liu}\ \emph {et~al.}(2022)\citenamefont {Liu},
  \citenamefont {Gao}, \citenamefont {Cain}, \citenamefont {Lukin},\ and\
  \citenamefont {Wang}}]{liu2022computing}%
  \BibitemOpen
  \bibfield  {author} {\bibinfo {author} {\bibfnamefont {J.-G.}\ \bibnamefont
  {Liu}}, \bibinfo {author} {\bibfnamefont {X.}~\bibnamefont {Gao}}, \bibinfo
  {author} {\bibfnamefont {M.}~\bibnamefont {Cain}}, \bibinfo {author}
  {\bibfnamefont {M.~D.}\ \bibnamefont {Lukin}}, \ and\ \bibinfo {author}
  {\bibfnamefont {S.-T.}\ \bibnamefont {Wang}},\ }\href@noop {} {\bibfield
  {journal} {\bibinfo  {journal} {arXiv preprint arXiv:2205.03718}\ } (\bibinfo
  {year} {2022})}\BibitemShut {NoStop}%
\bibitem [{\citenamefont {Wang}\ \emph {et~al.}(2023)\citenamefont {Wang},
  \citenamefont {Zhang}, \citenamefont {Pan},\ and\ \citenamefont
  {Zhang}}]{wang2023tensor}%
  \BibitemOpen
  \bibfield  {author} {\bibinfo {author} {\bibfnamefont {Y.}~\bibnamefont
  {Wang}}, \bibinfo {author} {\bibfnamefont {Y.~E.}\ \bibnamefont {Zhang}},
  \bibinfo {author} {\bibfnamefont {F.}~\bibnamefont {Pan}}, \ and\ \bibinfo
  {author} {\bibfnamefont {P.}~\bibnamefont {Zhang}},\ }\href@noop {}
  {\bibfield  {journal} {\bibinfo  {journal} {arXiv preprint arXiv:2305.01874}\
  } (\bibinfo {year} {2023})}\BibitemShut {NoStop}%
\bibitem [{\citenamefont {Fannes}\ \emph {et~al.}(1992)\citenamefont {Fannes},
  \citenamefont {Nachtergaele},\ and\ \citenamefont
  {Werner}}]{fannes1992ground}%
  \BibitemOpen
  \bibfield  {author} {\bibinfo {author} {\bibfnamefont {M.}~\bibnamefont
  {Fannes}}, \bibinfo {author} {\bibfnamefont {B.}~\bibnamefont
  {Nachtergaele}}, \ and\ \bibinfo {author} {\bibfnamefont {R.~F.}\
  \bibnamefont {Werner}},\ }\href@noop {} {\bibfield  {journal} {\bibinfo
  {journal} {Journal of statistical physics}\ }\textbf {\bibinfo {volume}
  {66}},\ \bibinfo {pages} {939} (\bibinfo {year} {1992})}\BibitemShut
  {NoStop}%
\bibitem [{\citenamefont {Friedman}(1997)}]{friedman1997density}%
  \BibitemOpen
  \bibfield  {author} {\bibinfo {author} {\bibfnamefont {B.}~\bibnamefont
  {Friedman}},\ }\href@noop {} {\bibfield  {journal} {\bibinfo  {journal}
  {Journal of Physics: Condensed Matter}\ }\textbf {\bibinfo {volume} {9}},\
  \bibinfo {pages} {9021} (\bibinfo {year} {1997})}\BibitemShut {NoStop}%
\bibitem [{\citenamefont {Lepetit}\ \emph {et~al.}(2000)\citenamefont
  {Lepetit}, \citenamefont {Cousy},\ and\ \citenamefont
  {Pastor}}]{lepetit2000density}%
  \BibitemOpen
  \bibfield  {author} {\bibinfo {author} {\bibfnamefont {M.-B.}\ \bibnamefont
  {Lepetit}}, \bibinfo {author} {\bibfnamefont {M.}~\bibnamefont {Cousy}}, \
  and\ \bibinfo {author} {\bibfnamefont {G.~M.}\ \bibnamefont {Pastor}},\
  }\href@noop {} {\bibfield  {journal} {\bibinfo  {journal} {The European
  Physical Journal B-Condensed Matter and Complex Systems}\ }\textbf {\bibinfo
  {volume} {13}},\ \bibinfo {pages} {421} (\bibinfo {year} {2000})}\BibitemShut
  {NoStop}%
\bibitem [{\citenamefont {Shi}\ \emph {et~al.}(2006)\citenamefont {Shi},
  \citenamefont {Duan},\ and\ \citenamefont {Vidal}}]{shi2006classical}%
  \BibitemOpen
  \bibfield  {author} {\bibinfo {author} {\bibfnamefont {Y.-Y.}\ \bibnamefont
  {Shi}}, \bibinfo {author} {\bibfnamefont {L.-M.}\ \bibnamefont {Duan}}, \
  and\ \bibinfo {author} {\bibfnamefont {G.}~\bibnamefont {Vidal}},\
  }\href@noop {} {\bibfield  {journal} {\bibinfo  {journal} {Physical review
  a}\ }\textbf {\bibinfo {volume} {74}},\ \bibinfo {pages} {022320} (\bibinfo
  {year} {2006})}\BibitemShut {NoStop}%
\bibitem [{\citenamefont {Tagliacozzo}\ \emph {et~al.}(2009)\citenamefont
  {Tagliacozzo}, \citenamefont {Evenbly},\ and\ \citenamefont
  {Vidal}}]{tagliacozzo2009simulation}%
  \BibitemOpen
  \bibfield  {author} {\bibinfo {author} {\bibfnamefont {L.}~\bibnamefont
  {Tagliacozzo}}, \bibinfo {author} {\bibfnamefont {G.}~\bibnamefont
  {Evenbly}}, \ and\ \bibinfo {author} {\bibfnamefont {G.}~\bibnamefont
  {Vidal}},\ }\href@noop {} {\bibfield  {journal} {\bibinfo  {journal}
  {Physical Review B}\ }\textbf {\bibinfo {volume} {80}},\ \bibinfo {pages}
  {235127} (\bibinfo {year} {2009})}\BibitemShut {NoStop}%
\bibitem [{\citenamefont {Nagaj}\ \emph {et~al.}(2008)\citenamefont {Nagaj},
  \citenamefont {Farhi}, \citenamefont {Goldstone}, \citenamefont {Shor},\ and\
  \citenamefont {Sylvester}}]{nagaj2008quantum}%
  \BibitemOpen
  \bibfield  {author} {\bibinfo {author} {\bibfnamefont {D.}~\bibnamefont
  {Nagaj}}, \bibinfo {author} {\bibfnamefont {E.}~\bibnamefont {Farhi}},
  \bibinfo {author} {\bibfnamefont {J.}~\bibnamefont {Goldstone}}, \bibinfo
  {author} {\bibfnamefont {P.}~\bibnamefont {Shor}}, \ and\ \bibinfo {author}
  {\bibfnamefont {I.}~\bibnamefont {Sylvester}},\ }\href@noop {} {\bibfield
  {journal} {\bibinfo  {journal} {Physical Review B}\ }\textbf {\bibinfo
  {volume} {77}},\ \bibinfo {pages} {214431} (\bibinfo {year}
  {2008})}\BibitemShut {NoStop}%
\bibitem [{\citenamefont {Murg}\ \emph {et~al.}(2010)\citenamefont {Murg},
  \citenamefont {Verstraete}, \citenamefont {Legeza},\ and\ \citenamefont
  {Noack}}]{murg2010simulating}%
  \BibitemOpen
  \bibfield  {author} {\bibinfo {author} {\bibfnamefont {V.}~\bibnamefont
  {Murg}}, \bibinfo {author} {\bibfnamefont {F.}~\bibnamefont {Verstraete}},
  \bibinfo {author} {\bibfnamefont {{\"O}.}~\bibnamefont {Legeza}}, \ and\
  \bibinfo {author} {\bibfnamefont {R.~M.}\ \bibnamefont {Noack}},\ }\href@noop
  {} {\bibfield  {journal} {\bibinfo  {journal} {Physical Review B}\ }\textbf
  {\bibinfo {volume} {82}},\ \bibinfo {pages} {205105} (\bibinfo {year}
  {2010})}\BibitemShut {NoStop}%
\bibitem [{\citenamefont {Li}\ \emph {et~al.}(2012)\citenamefont {Li},
  \citenamefont {von Delft},\ and\ \citenamefont {Xiang}}]{li2012efficient}%
  \BibitemOpen
  \bibfield  {author} {\bibinfo {author} {\bibfnamefont {W.}~\bibnamefont
  {Li}}, \bibinfo {author} {\bibfnamefont {J.}~\bibnamefont {von Delft}}, \
  and\ \bibinfo {author} {\bibfnamefont {T.}~\bibnamefont {Xiang}},\
  }\href@noop {} {\bibfield  {journal} {\bibinfo  {journal} {Physical Review
  B}\ }\textbf {\bibinfo {volume} {86}},\ \bibinfo {pages} {195137} (\bibinfo
  {year} {2012})}\BibitemShut {NoStop}%
\bibitem [{\citenamefont {Nakatani}\ and\ \citenamefont
  {Chan}(2013)}]{nakatani2013efficient}%
  \BibitemOpen
  \bibfield  {author} {\bibinfo {author} {\bibfnamefont {N.}~\bibnamefont
  {Nakatani}}\ and\ \bibinfo {author} {\bibfnamefont {G.~K.-L.}\ \bibnamefont
  {Chan}},\ }\href@noop {} {\bibfield  {journal} {\bibinfo  {journal} {The
  Journal of chemical physics}\ }\textbf {\bibinfo {volume} {138}},\ \bibinfo
  {pages} {134113} (\bibinfo {year} {2013})}\BibitemShut {NoStop}%
\bibitem [{\citenamefont {Pi{\v{z}}orn}\ \emph {et~al.}(2013)\citenamefont
  {Pi{\v{z}}orn}, \citenamefont {Verstraete},\ and\ \citenamefont
  {Konik}}]{pivzorn2013tree}%
  \BibitemOpen
  \bibfield  {author} {\bibinfo {author} {\bibfnamefont {I.}~\bibnamefont
  {Pi{\v{z}}orn}}, \bibinfo {author} {\bibfnamefont {F.}~\bibnamefont
  {Verstraete}}, \ and\ \bibinfo {author} {\bibfnamefont {R.~M.}\ \bibnamefont
  {Konik}},\ }\href@noop {} {\bibfield  {journal} {\bibinfo  {journal}
  {Physical Review B}\ }\textbf {\bibinfo {volume} {88}},\ \bibinfo {pages}
  {195102} (\bibinfo {year} {2013})}\BibitemShut {NoStop}%
\bibitem [{\citenamefont {Gerster}\ \emph {et~al.}(2014)\citenamefont
  {Gerster}, \citenamefont {Silvi}, \citenamefont {Rizzi}, \citenamefont
  {Fazio}, \citenamefont {Calarco},\ and\ \citenamefont
  {Montangero}}]{gerster2014unconstrained}%
  \BibitemOpen
  \bibfield  {author} {\bibinfo {author} {\bibfnamefont {M.}~\bibnamefont
  {Gerster}}, \bibinfo {author} {\bibfnamefont {P.}~\bibnamefont {Silvi}},
  \bibinfo {author} {\bibfnamefont {M.}~\bibnamefont {Rizzi}}, \bibinfo
  {author} {\bibfnamefont {R.}~\bibnamefont {Fazio}}, \bibinfo {author}
  {\bibfnamefont {T.}~\bibnamefont {Calarco}}, \ and\ \bibinfo {author}
  {\bibfnamefont {S.}~\bibnamefont {Montangero}},\ }\href@noop {} {\bibfield
  {journal} {\bibinfo  {journal} {Physical Review B}\ }\textbf {\bibinfo
  {volume} {90}},\ \bibinfo {pages} {125154} (\bibinfo {year}
  {2014})}\BibitemShut {NoStop}%
\bibitem [{\citenamefont {Murg}\ \emph {et~al.}(2015)\citenamefont {Murg},
  \citenamefont {Verstraete}, \citenamefont {Schneider}, \citenamefont {Nagy},\
  and\ \citenamefont {Legeza}}]{murg2015tree}%
  \BibitemOpen
  \bibfield  {author} {\bibinfo {author} {\bibfnamefont {V.}~\bibnamefont
  {Murg}}, \bibinfo {author} {\bibfnamefont {F.}~\bibnamefont {Verstraete}},
  \bibinfo {author} {\bibfnamefont {R.}~\bibnamefont {Schneider}}, \bibinfo
  {author} {\bibfnamefont {P.~R.}\ \bibnamefont {Nagy}}, \ and\ \bibinfo
  {author} {\bibfnamefont {O.}~\bibnamefont {Legeza}},\ }\href@noop {}
  {\bibfield  {journal} {\bibinfo  {journal} {Journal of Chemical Theory and
  Computation}\ }\textbf {\bibinfo {volume} {11}},\ \bibinfo {pages} {1027}
  (\bibinfo {year} {2015})}\BibitemShut {NoStop}%
\bibitem [{\citenamefont {Yedidia}\ \emph {et~al.}(2001)\citenamefont
  {Yedidia}, \citenamefont {Freeman},\ and\ \citenamefont
  {Weiss}}]{Yedidia2001}%
  \BibitemOpen
  \bibfield  {author} {\bibinfo {author} {\bibfnamefont {J.~S.}\ \bibnamefont
  {Yedidia}}, \bibinfo {author} {\bibfnamefont {W.}~\bibnamefont {Freeman}}, \
  and\ \bibinfo {author} {\bibfnamefont {Y.}~\bibnamefont {Weiss}},\ }in\ \href
  {https://proceedings.neurips.cc/paper/2000/file/61b1fb3f59e28c67f3925f3c79be81a1-Paper.pdf}
  {\emph {\bibinfo {booktitle} {Advances in Neural Information Processing
  Systems}}},\ Vol.~\bibinfo {volume} {13},\ \bibinfo {editor} {edited by\
  \bibinfo {editor} {\bibfnamefont {T.}~\bibnamefont {Leen}}, \bibinfo {editor}
  {\bibfnamefont {T.}~\bibnamefont {Dietterich}}, \ and\ \bibinfo {editor}
  {\bibfnamefont {V.}~\bibnamefont {Tresp}}}\ (\bibinfo  {publisher} {MIT
  Press},\ \bibinfo {year} {2001})\BibitemShut {NoStop}%
\bibitem [{Note1()}]{Note1}%
  \BibitemOpen
  \bibinfo {note} {The reader familiar with tensor networks, may a appreciate
  the connection with the general procedure of dealing with tensor networks
  with few loops. Namely, by cutting a given bond and storing the result as a
  sum over the corresponding singular values. Notice that, the sum grows
  exponentially with the numbers of cut bonds.}\BibitemShut {Stop}%
\bibitem [{\citenamefont {Ran}\ \emph {et~al.}(2020)\citenamefont {Ran},
  \citenamefont {Tirrito}, \citenamefont {Peng}, \citenamefont {Chen},
  \citenamefont {Tagliacozzo}, \citenamefont {Su},\ and\ \citenamefont
  {Lewenstein}}]{ran2020tensor}%
  \BibitemOpen
  \bibfield  {author} {\bibinfo {author} {\bibfnamefont {S.-J.}\ \bibnamefont
  {Ran}}, \bibinfo {author} {\bibfnamefont {E.}~\bibnamefont {Tirrito}},
  \bibinfo {author} {\bibfnamefont {C.}~\bibnamefont {Peng}}, \bibinfo {author}
  {\bibfnamefont {X.}~\bibnamefont {Chen}}, \bibinfo {author} {\bibfnamefont
  {L.}~\bibnamefont {Tagliacozzo}}, \bibinfo {author} {\bibfnamefont
  {G.}~\bibnamefont {Su}}, \ and\ \bibinfo {author} {\bibfnamefont
  {M.}~\bibnamefont {Lewenstein}},\ }\href@noop {} {\emph {\bibinfo {title}
  {Tensor Network Contractions: Methods and Applications to Quantum Many-Body
  Systems}}}\ (\bibinfo  {publisher} {Springer Nature},\ \bibinfo {year}
  {2020})\BibitemShut {NoStop}%
\bibitem [{\citenamefont {Pancotti}\ and\ \citenamefont
  {Gray}(2023)}]{gitrepo}%
  \BibitemOpen
  \bibfield  {author} {\bibinfo {author} {\bibfnamefont {N.}~\bibnamefont
  {Pancotti}}\ and\ \bibinfo {author} {\bibfnamefont {J.}~\bibnamefont
  {Gray}},\ }\href@noop {} {\enquote {\bibinfo {title} {{TNMPA: Tensor Network
  Message Passing Algorithms}},}\ }\bibinfo {howpublished}
  {\url{https://github.com/awslabs/tensor-message-passing}} (\bibinfo {year}
  {2023})\BibitemShut {NoStop}%
\bibitem [{\citenamefont {Gray}(2018)}]{gray2018quimb}%
  \BibitemOpen
  \bibfield  {author} {\bibinfo {author} {\bibfnamefont {J.}~\bibnamefont
  {Gray}},\ }\href@noop {} {\bibfield  {journal} {\bibinfo  {journal} {Journal
  of Open Source Software}\ }\textbf {\bibinfo {volume} {3}},\ \bibinfo {pages}
  {819} (\bibinfo {year} {2018})}\BibitemShut {NoStop}%
\bibitem [{\citenamefont {Pumphrey}(2001)}]{Pumphrey2001SOLVINGTS}%
  \BibitemOpen
  \bibfield  {author} {\bibinfo {author} {\bibfnamefont {S.}~\bibnamefont
  {Pumphrey}}\ }(\bibinfo {year} {2001})\BibitemShut {NoStop}%
\bibitem [{\citenamefont {Sharma}\ \emph {et~al.}(2019)\citenamefont {Sharma},
  \citenamefont {Roy}, \citenamefont {Soos},\ and\ \citenamefont
  {Meel}}]{SRSM19}%
  \BibitemOpen
  \bibfield  {author} {\bibinfo {author} {\bibfnamefont {S.}~\bibnamefont
  {Sharma}}, \bibinfo {author} {\bibfnamefont {S.}~\bibnamefont {Roy}},
  \bibinfo {author} {\bibfnamefont {M.}~\bibnamefont {Soos}}, \ and\ \bibinfo
  {author} {\bibfnamefont {K.~S.}\ \bibnamefont {Meel}},\ }in\ \href@noop {}
  {\emph {\bibinfo {booktitle} {Proceedings of International Joint Conference
  on Artificial Intelligence (IJCAI)}}}\ (\bibinfo {year} {2019})\BibitemShut
  {NoStop}%
\bibitem [{\citenamefont {Jiang}\ \emph {et~al.}(2008)\citenamefont {Jiang},
  \citenamefont {Weng},\ and\ \citenamefont {Xiang}}]{jiang2008accurate}%
  \BibitemOpen
  \bibfield  {author} {\bibinfo {author} {\bibfnamefont {H.-C.}\ \bibnamefont
  {Jiang}}, \bibinfo {author} {\bibfnamefont {Z.-Y.}\ \bibnamefont {Weng}}, \
  and\ \bibinfo {author} {\bibfnamefont {T.}~\bibnamefont {Xiang}},\
  }\href@noop {} {\bibfield  {journal} {\bibinfo  {journal} {Physical review
  letters}\ }\textbf {\bibinfo {volume} {101}},\ \bibinfo {pages} {090603}
  (\bibinfo {year} {2008})}\BibitemShut {NoStop}%
\bibitem [{\citenamefont {Sahu}\ and\ \citenamefont
  {Swingle}(2022)}]{sahu2022efficient}%
  \BibitemOpen
  \bibfield  {author} {\bibinfo {author} {\bibfnamefont {S.}~\bibnamefont
  {Sahu}}\ and\ \bibinfo {author} {\bibfnamefont {B.}~\bibnamefont {Swingle}},\
  }\href@noop {} {\bibfield  {journal} {\bibinfo  {journal} {arXiv preprint
  arXiv:2206.04701}\ } (\bibinfo {year} {2022})}\BibitemShut {NoStop}%
\bibitem [{\citenamefont {Cao}\ and\ \citenamefont {Vontobel}(2017)}]{Cao2017}%
  \BibitemOpen
  \bibfield  {author} {\bibinfo {author} {\bibfnamefont {M.~X.}\ \bibnamefont
  {Cao}}\ and\ \bibinfo {author} {\bibfnamefont {P.~O.}\ \bibnamefont
  {Vontobel}},\ }in\ \href {\doibase 10.1109/ITW.2017.8277985} {\emph {\bibinfo
  {booktitle} {2017 IEEE Information Theory Workshop (ITW)}}}\ (\bibinfo {year}
  {2017})\ pp.\ \bibinfo {pages} {136--140}\BibitemShut {NoStop}%
\bibitem [{\citenamefont {Laumann}\ \emph {et~al.}(2008)\citenamefont
  {Laumann}, \citenamefont {Scardicchio},\ and\ \citenamefont
  {Sondhi}}]{laumann2008cavity}%
  \BibitemOpen
  \bibfield  {author} {\bibinfo {author} {\bibfnamefont {C.}~\bibnamefont
  {Laumann}}, \bibinfo {author} {\bibfnamefont {A.}~\bibnamefont
  {Scardicchio}}, \ and\ \bibinfo {author} {\bibfnamefont {S.}~\bibnamefont
  {Sondhi}},\ }\href@noop {} {\bibfield  {journal} {\bibinfo  {journal}
  {Physical Review B}\ }\textbf {\bibinfo {volume} {78}},\ \bibinfo {pages}
  {134424} (\bibinfo {year} {2008})}\BibitemShut {NoStop}%
\bibitem [{\citenamefont {Leifer}\ and\ \citenamefont
  {Poulin}(2008)}]{leifer2008quantum}%
  \BibitemOpen
  \bibfield  {author} {\bibinfo {author} {\bibfnamefont {M.~S.}\ \bibnamefont
  {Leifer}}\ and\ \bibinfo {author} {\bibfnamefont {D.}~\bibnamefont
  {Poulin}},\ }\href@noop {} {\bibfield  {journal} {\bibinfo  {journal} {Annals
  of Physics}\ }\textbf {\bibinfo {volume} {323}},\ \bibinfo {pages} {1899}
  (\bibinfo {year} {2008})}\BibitemShut {NoStop}%
\end{thebibliography}%

\appendix
\onecolumngrid

\section{Numerical Implementation}\label{appdx:algorithms}

In this section, we provide the details of the algorithms above, and the data collected in Fig.~\ref{fig:solutions}.
Algorithm~\ref{alg:tnbp} describes the core message passing routines that are used to optimize the environments.
They apply both in the case of belief propagation and survey propagation.
In practice, the only difference between BP and SP is the update procedure in line seven and nine.
And they differ on the explicit form for the tensors that were given in the main text.
The data in Fig.~\ref{fig:solutions} was generated with $\delta = 0.001$ and $m = 1000$.

Algorithm~\ref{alg:tnsp} finds a solution of a given {\it k}-SAT instance by using the message passing procedure in Algorithm~\ref{alg:tnbp}.
It iteratively computes the approximate fixed-point environments.
That are used to find the variable with the largest absolute bias.
Variables are fixed according to the sign of the bias one after the other as long as the absolute value of the bias is larger than a certain threshold.
This is not strictly necessary for BP.
The algorithm can succeed even with very small biases.
In the case of SP, we decimate the instance until TNBP can only find trivial fixed points, i.e. small absolute biases.
We can finally solve the decimated instance with simpler algorithms like WALKSAT.

\begin{algorithm}[t]
\begin{algorithmic}[1]
  \REQUIRE
  Collection of tensors: $\{T_j, T_a\}$, Tolerance: $\delta$, Max number of iterations: $m$\\
  \STATE For each variable $i$ initialize $|\partial i|$ environment vectors and
  \STATE For each clause $a$ initialize $|\partial a|$ environments vectors
  \STATE $d = 0$
  \FOR{$t \in \{ 1, \ldots m \}$}
  \FOR{\textbf{each} $i \in V$}
  \FOR{\textbf{each} $a \in |\partial i|$}
  \STATE   $E^{(t+1)}_{j \rightarrow a} [r_{j_a}] = \sum_{r_{\partial j \backslash j_a}} T_j [r_{\partial j}] \bigotimes_{b \in \partial j \backslash a} E^{(t)}_{b \rightarrow j} [r_{b_j}]$
  \STATE $d = \max(|E^{(t + 1)}_{i \rightarrow a} - E^{(t)}_{i \rightarrow a}|, d)$
  \STATE $E^{(t+1)}_{a \rightarrow j} [r_{a_j}] = \sum_{r_{\partial a \backslash j_a}} T_a [r_{\partial a}]  \bigotimes_{k \in \partial a \backslash j} E^{(t)}_{k \rightarrow a} [r_{a_k}]$
  \STATE $d = \max(|E^{(t + 1)}_{a \rightarrow i} - E^{(t)}_{a \rightarrow i}|, d)$
  \ENDFOR
  \ENDFOR
  \IF{$d<\delta$}
  \STATE Return SUCCESS, $\{E^{(t + 1)}_{a \rightarrow i}, E^{(t + 1)}_{i \rightarrow a} \}$
  \ENDIF
  \ENDFOR
  \STATE Return FAILURE, $\{E^{(t + 1)}_{a \rightarrow i}, E^{(t + 1)}_{i \rightarrow a} \}$
\end{algorithmic}
\caption{TNBP}
\label{alg:tnbp}
\end{algorithm}

\begin{algorithm}[t]
\begin{algorithmic}[1]
  \REQUIRE
  {\it k}-SAT Instance: $\mathcal{I}(N, M)$, Tolerance: $\delta$, Max number of iterations: $m$\\
  \STATE Create the collection of tensors $\{T_j, T_a\}$ that represent $\mathcal{I}$ (BP or SP)
  \WHILE{True}
  \STATE STATUS, $\{E^*_{a \rightarrow i}, E^*_{i \rightarrow a} \}$  = TNBP($\{T_j, T_a\}$, $\delta$, $m$).
  \IF {STATUS = FAILURE}
  \STATE return FAILURE
  \ENDIF
  \FOR{\textbf{each} $i \in V$}
  \STATE Compute biases $b_i(\{E^*_{a \rightarrow i}\}_{a \in \partial i})$
  \ENDFOR
  \STATE $B = \max_i(b_i)$ and $v = \arg\max_i(b_i)$
  \IF{$|B| < \delta$}
  \STATE Call WALKSAT($\mathcal{I}$, $f$, $p$)
  \ELSE
  \STATE Fix $v$ according to the sign of $B$
  \STATE Update the instance $\mathcal{I}$: remove $v$ and the clauses satisfied by that
  \ENDIF
  \ENDWHILE
\end{algorithmic}
\caption{Decimation (BP, SP)}
\label{alg:tnsp}
\end{algorithm}

\begin{algorithm}[t]
\begin{algorithmic}[1]
  \REQUIRE
  {\it k}-SAT Instance: $\mathcal{I}(N, M)$, number of flips: $f$, mixing: $p$\\
  \STATE $t = 0$
  \STATE Initialize $X$ to a random configuration
  \WHILE{$t < f$}
  \IF {$E(X) = 0$}
  \STATE Return SUCCESS
  \ENDIF
  \STATE Let $r$ be uniformly random in $[0, 1]$
  \IF {$r < 1-p$}
  \FOR {\textbf{each} $i \in V$}
  \STATE $\Delta_i = E(X(i)) - E(X)$. ($X(i)$ is equal to $X$, with $X_i$ is flipped)
  \ENDFOR
  \STATE $v = \arg\min_i(\Delta_i)$
  \STATE Flip $v$
  \ELSE
  \STATE Choose a violated clause $a$ at random
  \STATE Pick at random a variable $X_j$ with $j \in \partial a$ and flip it.
  \ENDIF
  \ENDWHILE
  \STATE Return FAILURE
\end{algorithmic}
\caption{WALKSAT}
\label{alg:walksat}
\end{algorithm}

\section{Free entropy in terms of environments}

The Bethe free entropy can be written as the sum of internal energy and entropy by using Eq.~\eqref{eq:joint_dist_envs} as
\begin{equation}
    \mathbf{F}_T [m] = - \sum_{i} \sum_{x_i, r_{\partial i}} m_i (x_i, r _{\partial i}) \log \frac{m_i (x_i, r _{\partial i})} {T_i (x_i, r_ {\partial i})} + \sum_{j} \sum_{r_{\partial i}} m_j (r_j) \log m_j (r_j).
\end{equation}
If we write the marginals in terms of environments, we get for the first term
\begin{align*}
 - \sum_{i \in V} \sum_{x_i, r_{\partial i}} m_i (x_i, r _{\partial i}) \log \frac{m_i (x_i, r _{\partial i})} {T_i (x_i, r_ {\partial i})} &= - \sum_{i \in V} \sum_{x_i, r_{\partial i}} m_i (x_i, r _{\partial i}) \log \frac{T_i (x_i, r_{\partial i}) \prod_{j \in \partial i } E^*_{j \rightarrow i} (r_b)} {Z_{\partial i}T_i (x_i, r_ {\partial i})} \\
  &= \sum_{i \in V} \log Z_{\partial i} (E_{\partial i}) - \sum_{i \in V} \sum_{r_b} \sum_{x_i, r_c \backslash r_b} m_i (x_i, r _{\partial i}) \sum_{j \in \partial i } \log E^*_{j \rightarrow i} (r_b) \\
  &= \sum_{i \in V} \log Z_{\partial i} (E_{\partial i}) - \sum_{i \in V} \sum_{j \in \partial i } \sum_{r_b} m_i (r _b) \log E^*_{j \rightarrow i} (r_b) \\
  &= \sum_{i \in V} \log Z_{\partial i} (E_{\partial i}) - \sum_{i \in V_r} \sum_{r_b} m_i (r _b) \log E^*_{j \rightarrow i} (r_b) E^*_{i \rightarrow j} (r_b),
\end{align*}
and for the second term
\begin{align*}
  \sum_{j \in V_r} \sum_{r_{\partial i}} m_j (r_j) \log m_j (r_j) &= \sum_{j \in V_r} \sum_{r_{\partial i}} m_j (r_j) \log \frac{E^*_{j \rightarrow i} (r_j) E^*_{i \rightarrow j} (r_j)}{Z_{ij}} \\
  &= - \sum_{j \in V_r} \log Z_{ij} (E_{i\rightarrow j}, E_{j \rightarrow i}) + \sum_{j \in V_r} \sum_{r_{\partial i}} m_j (r_j) \log E^*_{j \rightarrow i} (r_j) E^*_{i \rightarrow j} (r_j).
\end{align*}

Thus, the total free entropy can be written in terms of ``local partition functions'' as 

\begin{equation*}
  \mathbf{F}_T [E] = \sum_{i \in V} \log Z_{\partial i} (E_{\partial i}) - \sum_{j \in V_r} \log Z_{ij} (E_{i\rightarrow j}, E_{j \rightarrow i}) \\
\end{equation*}

A similar argument holds in the quantum case. 
For more details, see Sec.~\ref{sec:quantum}.

\end{document}